\documentclass[useAMS,usegraphicx]{mn2e}
\usepackage{rotate}
\usepackage{times}
\newif\ifAMStwofonts
\AMStwofontstrue

\newcommand{\me}{\mathrm{e}}
\newcommand{\mpi}{\mathrm{\pi}}
\newcommand{\mi}{\mathrm{i}}

\def\gs{\mathrel{\hbox{\rlap{\hbox{\lower4pt\hbox{$\sim$}}}\hbox{$>$}}}}
\def\ls{\mathrel{\hbox{\rlap{\hbox{\lower4pt\hbox{$\sim$}}}\hbox{$<$}}}}

\def\Msun{M$_{\odot}$}

\def\ginga{{\it Ginga}}
\def\exosat{{\it EXOSAT}}
\def\xmm{{\it XMM-Newton}}

\def\asca{{\it ASCA}}
\def\sax{{\it BeppoSAX}}
\def\xte{{\it RXTE}}
\def\xmm{{\it XMM-Newton}}
\def\conx{{\it Constellation-X}}
\def\xeus{{\it XEUS}}

\def\et{{et al.\ }}
\def\mcg{{MCG--6-30-15}}

\def\Msun{\hbox{$\rm\thinspace M_{\odot}$}}

\title[Variability of \mcg]
      {X-ray continuum variability of \mcg}

\author[Vaughan \et]
       {S. Vaughan,$^{1}$\thanks{E-mail: sav@ast.cam.ac.uk}
        A. C. Fabian$^{1}$ and
        K. Nandra$^{2}$\\
$^{1}$Institute of Astronomy, University of Cambridge, Madingley Road, Cambridge CB3 0HA\\
$^{2}$Laboratory for High Energy Astrophysics,  NASA/Goddard Space Flight Center, Greenbelt, MD 20771, USA\\
}
\date{Accepted 19/11/2002; submitted 12/11/2002; in original form 1/10/2002}

\pagerange{\pageref{firstpage}--\pageref{lastpage}}
\pubyear{2002}

\begin{document}
\maketitle
\label{firstpage}

\begin{abstract}
This paper presents a comprehensive examination of the X-ray continuum
variability of the bright Seyfert 1 galaxy \mcg.  The source clearly
shows the strong, linear correlation between rms variability amplitude
and flux first seen in Galactic X-ray binaries.  The high frequency
power spectral density (PSD) of \mcg\ is examined in detail using a
Monte Carlo fitting procedure and is found to be well represented by a
steep power-law at high frequencies (with a power-law index $\alpha
\approx 2.5$), breaking to a flatter slope ($\alpha \approx 1$) below
$f_{\rm br} \approx 0.6 - 2.0 \times 10^{-4}$~Hz, consistent
with the previous results of Uttley, M$^{\rm c}$Hardy \& Papadakis. The slope
of the power spectrum above the break is energy dependent, with the
higher energies showing a flatter PSD. At low frequencies the
variations between different energy bands are highly coherent while at
high frequencies the coherence is significantly reduced. Time lags are
detected between energy bands, with the soft variations leading the
hard. The magnitude of the lag is small ($\ls 200$~s for the frequencies
observed) and is most likely frequency dependent. 
These properties are remarkably similar to the temporal properties of
the Galactic black hole candidate Cygnus X-1. The
characteristic timescales in these two types of source differ by
$\sim 10^5$; assuming that these timescales scale linearly with black hole
mass then suggests a black hole mass $\sim 10^6$~\Msun\ for \mcg. We
speculate that the timing properties of \mcg\ may be analogous to
those of Cyg X-1 in its high/soft state and discuss a simple phenomenological model,
originally developed to explain the timing properties of Cyg X-1, that
can explain many of the observed properties of \mcg.
\end{abstract}

\begin{keywords}
galaxies: active -- galaxies: Seyfert: general -- galaxies:
individual: \mcg\ -- X-ray: galaxies 
\end{keywords}

\section{Introduction}

The X-ray emission from radio-quiet Active Galactic Nuclei (AGN) shows
persistent  
variability  that is stronger and more rapid than in any other
waveband. This rapid X-ray variability provided early support for the
black hole/accretion disc model of AGN (e.g. Rees 1984) and suggested
that the X-ray emission is produced in the inner regions of the
central engine (see Mushotzky, Done \& Pounds 1993 for a review).  The
X-ray fluctuations appear erratic and random, with variability
occurring on a range of timescales (e.g. M$^{\rm c}$Hardy 1989).

In the study of time-variable processes one of the most widely used
statistical tools is the variability power spectrum (e.g. Priestley
1981; Bloomfield 2000).  The power spectral density (PSD) describes
the amount of variability power (i.e. amplitude$^{2}$) as a function
of temporal frequency (timescale$^{-1}$).  The long, uninterrupted
observations provided by \exosat\ allowed for the measurement of the
first accurate X-ray PSDs for Seyfert 1 galaxies.  These \exosat\ PSDs were
shown to rise at lower frequencies as a power-law: $\mathcal{P}(f)
\propto f^{-\alpha}$, where $\mathcal{P}(f)$ is the power at frequency
$f$ and $\alpha$ is the PSD slope, found to be $\alpha
\sim 1.5$ over the frequencies $\sim 10^{-5}$ to $\sim 10^{-3}$~Hz
(Green, M$^{\rm c}$Hardy \& Lehto 1993; Lawrence \& Papadakis 1993).  This
type of broad-band variability, rising to lower frequencies with a
power spectral slope $\alpha \gs 1$, is usually called ``red noise''
(for an introduction to red noise in astronomy and elsewhere see Press
1978).

The red noise PSDs of Seyfert 1s are similar to those observed in
Galactic Black Hole Candidates (GBHCs, see e.g. van der Klis 1995) on much
shorter timescales, and suggests the X-ray emission mechanisms may be
the same in these objects that differ in black hole mass by factors of
$\gs 10^{5}$ (M$^{\rm c}$Hardy 1989). The PSD of the best studied
GBHC, Cygnus X-1 in its 
low/hard state, can be approximated by a doubly-broken power-law with a
steep slope at high frequencies ($\alpha \approx 2$ above $\sim
3$~Hz), breaking to a flatter slope at intermediate frequencies
($\alpha \approx 1$ in the range $\approx 0.2-3$~Hz) then breaking to
a flat (``white'') spectrum at low frequencies ($\alpha \approx 0$
below $0.2$~Hz).  See Nowak \et (1999a) and Belloni \& Hasinger (1990)
for more details of Cyg X-1.  
Other GBHCs such as GX 339--4 show similar behaviour (Nowak, Wilms \&
Dove 2002). The position of the breaks in the PSD
represent ``characteristic timescales'' of the system, but 
these timescales have no clear physical interpretation at present
(e.g. Belloni, Psaltis \& van der Klis 2002 and references therein).

\begin{figure*}
\rotatebox{270}{
\scalebox{0.75}{\includegraphics{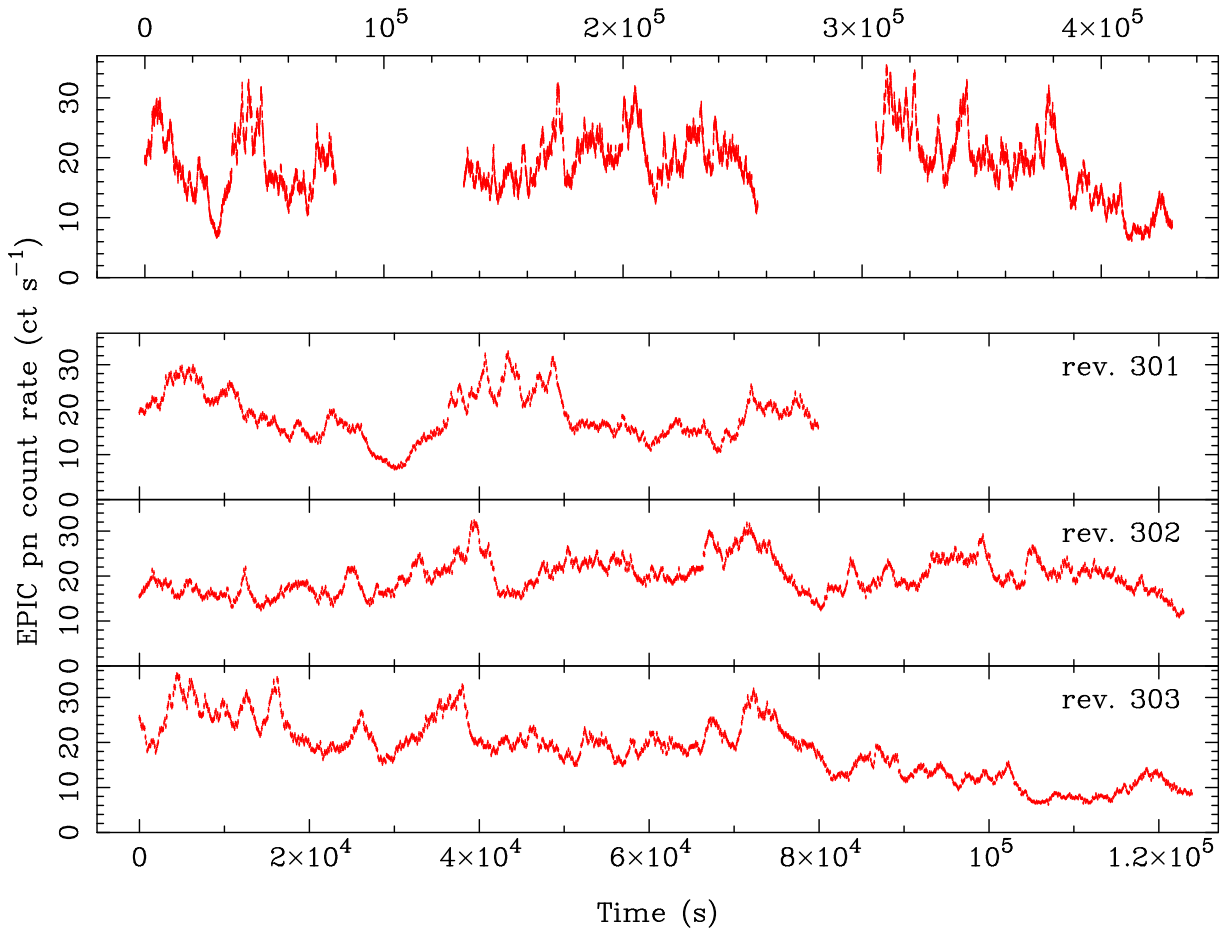}}}
\caption{
Full-band (0.2--10.0~keV) EPIC pn light curves in 100~s bins. 
The top panel shows all three consecutive light curves
and the bottom panels show the individual light curves from each \xmm\
revolution.
}
\label{fig:lightcurves}
\end{figure*}

The
featureless \exosat\ PSDs of Seyfert 1s provided no such timescales.
Such breaks must exist, however, or the total power would diverge at
low frequencies (for PSD slopes $\alpha \ge 1$).
Indeed, in recent years dedicated X-ray monitoring programmes with
\xte\ have revealed deviations from the power-law PSD in a few Seyfert
1 galaxies; at low frequencies the PSD breaks to a flatter slope
(Edelson \& Nandra 1999; Uttley, M$^{\rm c}$Hardy \& Papadakis 2002;
Markowitz \et 2002). The
timescale of this break frequency is consistent with the idea that the
PSD of AGN and GBHCs are essentially the same, with the
characteristic timescales scaling linearly with the black hole mass. 
Such a linear scaling of timescales with black hole
mass might be expected if the characteristic timescales are related to
e.g. a light-crossing timescale or an orbital timescale.
This linear scaling also means that a 100~ks light
curve of a Seyfert galaxy (such as presented here) is comparable to a $\ls
1$~s light curve of Cyg X-1. On these timescales a bright Seyfert 1
galaxy provides many more photons (by more than two orders of
magnitude) than the GBHC, i.e. Seyferts have a higher count rate per
characteristic timescale.  This means that intensive monitoring of
Seyferts can provide a view of the very high frequency variations in
accreting black holes that is difficult to access even in bright
GBHCs.

Not only do AGN and GBHCs show similar PSDs, but they display other
striking similarities in their X-ray timing properties. Uttley \&
M$^{\rm c}$Hardy (2001) showed that both X-ray binary and AGN light curves show
a tight, linear correlation between flux and rms variability amplitude
(see also Edelson \et 2002). The variability properties of GBHCs such
as Cyg X-1 are a function of photon energy. The PSD shape above the high
frequency break is energy-dependent, with the harder energy bands
showing flatter PSD slopes (e.g. Nowak \et 1999a; Lin \et 2000). In addition the
variations between different bands are highly coherent (well
correlated at each Fourier frequency) and show time delays, with
variations at softer energies leading those at harder energies
(Miyamoto \& Kitamoto 1989; Cui \et 1997; Nowak \et 1999a).
The magnitude of the time delay is frequency dependent, with the delay
increasing at lower frequencies. To date these properties have only
been measured in one AGN, NGC 7469 (Nandra \& Papadakis 2001;
Papadakis, Nandra \& Kazanas 2001), and in this object the
energy-dependence of the PSD slope, the coherence and time delay
properties appeared remarkably similar to those seen in Cyg X-1.
These similarities re-enforce the idea that the
X-ray emission mechanisms are the same in GBHCs and AGN.

This paper presents a detailed analysis of the X-ray continuum
variability of the bright, variable Seyfert 1 galaxy \mcg\
($z=0.007749$) using a long \xmm\ observation. The large collecting
area, wide band-pass and long orbit ($\sim 2$ day) of \xmm\ make it
ideal for examining the high frequency ($\gs 10^{-5}$~Hz) timing
properties of Seyfert galaxies. The analysis concentrates on frequency
domain (i.e. Fourier) time series analysis techniques, as these are
used extensively in the analysis of GBHC data.  The rest of this
paper is organised as follows. In Section~\ref{sect:data} the data
reduction procedures are outlined. In Section~\ref{sect:timing} the
basic properties of the light curves are discussed, followed by an
analysis of the PSD in Section~\ref{sect:pds}. Spectral variability is
then examined in detail by measuring the coherence and time lags
between different energy bands, as discussed in
Section~\ref{sect:cross}.  As
these analyses are relatively new to AGN studies we discuss the
methodology in some detail. 
Finally, the implications of these
results are discussed in Section~\ref{sect:disco} and some concluding
remarks are given in Section~\ref{sect:conc}.

\section{Data Reduction}
\label{sect:data}

\xmm\ (Jansen \et 2001) observed \mcg\ over the period 2001 July 31 --
2001 August 5 (revolutions 301, 302 and 303), during which all
instruments were operating nominally. Fabian \et (2002) discuss
details of the observation and basic data reduction. For the present
paper the data were reprocessed entirely with the latest software
({\tt SAS v5.3.3}) but the procedure followed that discussed in the
previous paper (except that the extraction region was a circle of radius
35 arcsec for both MOS and pn). 

Light curves were extracted from the EPIC pn data in four different
energy bands: 0.2--10.0~keV (full band), 0.2-0.7~keV (soft band),
0.7-2.0~keV (medium band) and 2.0-10.0~keV (hard band). These were
corrected for telemetry drop outs (less than 1 per cent of the total
time), background subtracted and binned to 100~s time resolution. The
errors on the light curves were calculated by propagating the Poisson
noise.  The light curves were not corrected for the $\sim 71$ per cent
``live time'' of the pn camera (Str\"{u}der \et 2001), which is only a
scaling factor. The EPIC MOS light curves were also examined, but
these are essentially identical to the pn light curves except with
lower signal-to-noise. Therefore, in the rest of this paper the
analysis concentrates on the pn data, but the MOS data are used to
confirm the pn results of Section~\ref{sect:cross}. 
The full band light curves are shown in
Fig.~\ref{fig:lightcurves}. The revolutions provided three
uninterrupted light curves of 80, 123 and 124~ks duration.  

As
discussed in Fabian \et (2002), \sax\ observed \mcg\ simultaneously
with \xmm, but these data are not used in the present paper. The LECS and
MECS data cover approximately the same bandpass as the EPIC (but with
lower signal to noise) and the higher energy PDS data (with a count
rate $\sim 0.7$ ct s$^{-1}$) do not have sufficient statistics to
determine the high energy variability properties accurately. In
addition, analysis of the \sax\ data is  complicated by the  poor
sampling of the light curves (due to repeated Earth occulatations and
SAA passages), indeed the amount of ``good''  exposure time obtained
for the PDS was $40$~ksec spread over a $>400$~ksec observing period.

\section{Basic temporal properties}
\label{sect:timing}

The source shows a factor of five variation between minimum and
maximum flux (both occurring during revolution 303), and significant
variability can be seen on timescales as short as $\sim 100$~s. 

As a first measure of the variability amplitude the root mean square (rms)
amplitude, in excess of Poisson noise, was calculated as 
\begin{equation}
\label{eqn:fvar}
\sigma_{\rm rms} = \sqrt{       S^{2} - \overline{\sigma_{\rm{err}}^{2}}},
\end{equation}
where $S^{2}$ is the total variance of the light curve $x(t_{i})$ of
length $N$ data points 
\begin{equation}
\label{eqn:variance}
S^{2} = \frac{1}{N-1} \sum_{i=1}^{N} (x(t_{i}) - \bar{x})^{2},
\end{equation}
where $\bar{x}$ is the mean of $x(t_{i})$
and $\overline{\sigma_{\rm{err}}^{2}}$ is the contribution expected
from measurement errors 
\begin{equation}
\label{eqn:mean_error}
\overline{\sigma_{\rm{err}}^{2}} = \frac{1}{N}\sum_{i=1}^{N} \left[ \sigma_{{\rm
err}}(t_i) \right]^{2},
\end{equation}
and $\sigma_{{\rm err}}(t_i)$ is the error on $x(t_{i})$.

The fractional excess rms variability amplitude ($F_{\rm var} =
\sigma_{\rm rms}/\bar{x}$; Edelson \et 2002) and mean count rate ($\bar{x}$) were
calculated for each of the three revolutions and are tabulated in
Table~\ref{tab:revs}. The errors on $F_{\rm var}$ were estimated
using the prescription given by Edelson \et (2002; but see their appendix
A for caveats on the interpretation of these estimates).
The medium band showed the highest variability amplitude. This can also
been seen in Fig.~6 of Fabian \et
(2002), where the fractional rms spectrum peaked in the 0.7--2.0~keV band.

\begin{table}
\centering
\caption{
Mean count rates ($\bar{x}$) and fractional excess rms variability amplitudes 
($F_{\rm var}$) for the light curves from each revolution and in each
energy band.
\label{tab:revs}} 
\begin{center}
\begin{tabular}{lccc}                
\hline
Revolution     & band & $\bar{x}$ (ct s$^{-1}$) & $F_{\rm var}$ (\%) \\
\hline
301            & Full   & $18.51$ & $28.7\pm0.7$ \\
               & Soft   & $ 8.63$ & $27.8\pm0.7$ \\
               & Medium & $ 6.68$ & $32.3\pm0.8$ \\
               & Hard   & $ 3.27$ & $26.0\pm0.7$ \\
\hline
302            & Full   & $19.94$ & $19.3\pm0.4$ \\
               & Soft   & $ 8.85$ & $18.4\pm0.4$ \\
               & Medium & $ 7.53$ & $21.4\pm0.4$ \\
               & Hard   & $ 3.61$ & $19.3\pm0.4$ \\
\hline
303            & Full   & $18.41$ & $34.9\pm0.7$ \\
               & Soft   & $ 8.56$ & $31.8\pm0.6$ \\
               & Medium & $ 6.69$ & $40.1\pm0.8$ \\
               & Hard   & $ 3.22$ & $33.7\pm0.7$ \\
\hline
\end{tabular}
\end{center}
\end{table}

\subsection{Radiative efficiency}

\begin{figure}
\rotatebox{270}{
\resizebox{!}{\columnwidth}{\includegraphics{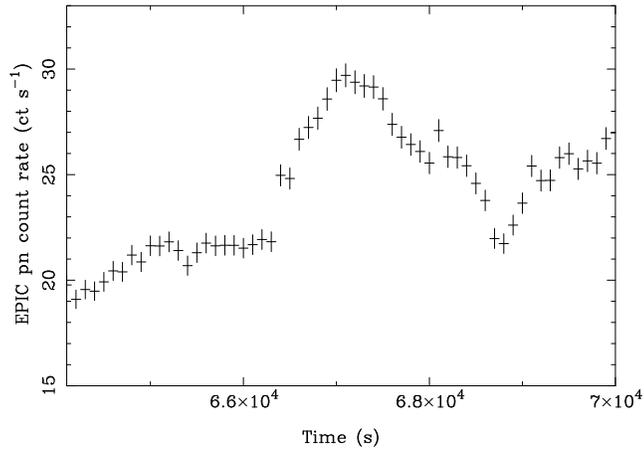}}}
\caption{
Section of the full-band (0.2--10.0~keV) EPIC pn light curve 
(from rev. 302) showing a rapid ``flare'' event.
}
\label{fig:flare}
\end{figure}

Fig.~\ref{fig:flare} shows a rapid ``event'' from the revolution 302
light curve. During this the source changed in 0.2--10~keV luminosity by $\Delta L
\approx 1.5 \times 10^{42}$~erg
s$^{-1}$ in 100~s. This was used to estimate the radiative efficiency
$\eta$ of  the source, assuming photon diffusion through a spherical
mass of accreting matter: $ \Delta L / \Delta t \ls \eta \cdot 2.1
\times 10^{42}$~erg s$^{-2}$ (Fabian 1979; Brandt \et 1999).  This gave a limit of
$\eta \gs 0.7$ per cent, below the expected maximum for accretion onto
a black hole (6 and 30 per cent for non-rotating and maximally rotating
black holes, respectively; Thorne 1974).

\subsection{rms-flux correlation}
\label{sect:rms-flux}

\begin{figure}
\rotatebox{270}{
\resizebox{!}{\columnwidth}{\includegraphics{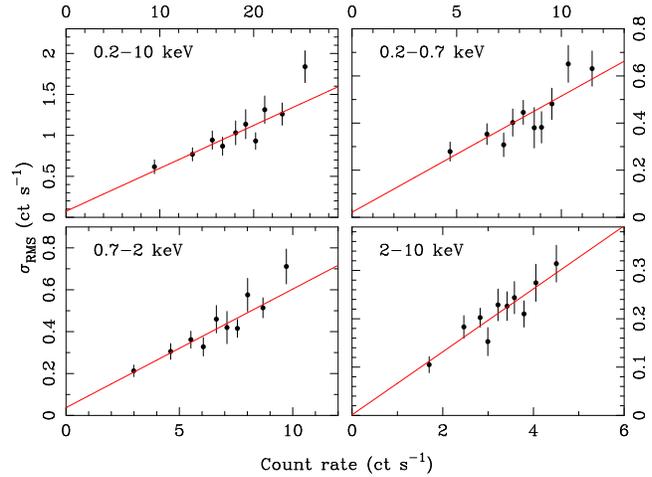}}}
\caption{
Correlation between rms variability amplitude and flux.
}
\label{fig:rmsflux}
\end{figure}

Using the 100~s resolution light curves the mean count rate and
$\sigma_{\rm rms}$ were calculated in bins of 15 data points. The
values of $\sigma_{\rm rms}$ from all three orbits were then binned as
a function of count rate such that there were 20 measurements per bin,
and an error was assigned based on the scatter within each bin
(equation 4.14 of Bevington \& Robinson 1992).  Fig.~\ref{fig:rmsflux}
shows the resulting correlation between $\sigma_{\rm rms}$ and count
rate in the four energy bands. The significance of the correlation was
tested using the Spearman rank-order correlation coefficient, $r_s$,
and the Kendall $\tau$ coefficient (Press \et 1992). Both these
indicate that the correlation is significant at $\gs 99.5$ per cent
confidence in all energy bands.  The effect of the rms-flux
correlation can be seen in the light curves
(Fig.~\ref{fig:lightcurves}): the peaks of the light curves appear
more variable (``jagged'') while the troughs appear much smoother.

Table~\ref{tab:rmsflux} shows the results of fitting a linear model to
these data; in all cases the linear model gives a good fit to the
data, with a $y$-axis intercept consistent with zero.  The gradient of
this relation corresponds to the $F_{\rm var}$ measured only on these
timescales (between 100 and 1500~s). The $1\sigma$ upper limit on any
constant flux offset ($x$-intercept) is $1.2$~ct s$^{-1}$ for the full
band, or $6.2$ per cent of the total count rate in that band. In the
three energy sub-bands the upper limits on the constant flux offset
are as follows: $1.04$~ct s$^{-1}$ ($12$ per cent), $0.09$~ct s$^{-1}$
($1.3$ per cent) and $0.43$~ct s$^{-1}$ ($12.8$ per cent) for the
soft, medium and hard bands respectively.

\begin{table}
\centering
\caption{Results of fitting a linear function to the 
rms-flux correlation. Errors on the model parameters correspond
  to a 90 per cent confidence level for one interesting parameter
  (i.e. a $\Delta \chi^{2}=2.7$ criterion).\label{tab:rmsflux}} 
\begin{center}
\begin{tabular}{lccc}                
\hline
Band      & $y$-intercept & Gradient & $\chi^{2}/dof$ \\
\hline
Full & $0.07\pm0.22$ & $0.052\pm0.013$ & $10.66/10$ \\
Soft & $0.02\pm0.12$ & $0.049\pm0.015$ & $7.78/10$ \\
Medium & $ 0.04\pm0.07$ & $0.057\pm0.016$ & $6.97/10$ \\
Hard   & $0.001\pm0.049$ & $0.065\pm0.016$ & $6.68/10$ \\
\hline
\end{tabular}
\end{center}
\end{table}

\section{Power spectral properties}
\label{sect:pds}

\subsection{Estimating the power spectrum}
\label{sect:pds_intro}

The PSD (sometimes known as the auto spectrum) describes the
amount of variability ``power'' as a function of Fourier
frequency. The PSD and auto correlation function (ACF) are Fourier
pairs, i.e. they are Fourier transforms of one another.
For a general introduction to Fourier methods of time series
analysis see e.g. Priestley (1981) and Bloomfield (2000), and for a
review of Fourier techniques as applied to X-ray time series analysis
see van der Klis (1989).

\subsubsection{Measuring the periodogram}
\label{sect:periodogram}

If the light curve $x(t_i)$  is evenly sampled (with a sampling period
$\Delta T$) then the PSD can be estimated by calculating the
periodogram\footnote{ Following Priestley (1981) we use the term
``periodogram'' to denote the discrete function $P(f_{j})$ described
by equation~\ref{eqn:pds}, which is merely an estimator of the
continuous PSD $\mathcal{P}(f)$. The periodogram is therefore related
to each new realisation of the process, whereas the PSD is
representative of the true, underlying process.}. This is the
normalised, modulus-squared of the Discrete Fourier Transform (DFT) of
the data. The DFT  of the light curve, $X(f_j)$, is calculated
as (see Press \et 1992):
\begin{equation}
\label{eqn:ft}
X(f_{j}) = 
\left( \frac{\Delta T}{N} \right)^{1/2} \sum_{i=1}^{N} x(t_i)~\me^{2\mpi \mi f_{j} t_{i}} =
R_X(f_j) + \mi I_X(f_j),
\end{equation}
where $R_X(f_j)$ and $I_X(f_j)$ are the real and imaginary parts of
the DFT given by the discrete cosine and sine transforms,
respectively:
\begin{eqnarray}
\label{eqn:sine}
R_X(f_j) = \left( \frac{\Delta T}{N} \right)^{1/2} \sum_{i=1}^{N} x(t_i)~ \cos (2 \mpi f_{j} t_{i})
\nonumber \\
I_X(f_j) = \left( \frac{\Delta T}{N} \right)^{1/2} \sum_{i=1}^{N} x(t_i)~ \sin (2 \mpi f_{j} t_{i}).
\end{eqnarray}

The transforms are normalised by $(\Delta T/N)^{1/2}$ so
that the amplitudes are independent of the sampling and length of the
light curve.
The DFT is calculated at $N/2$ evenly spaced frequencies  $f_{j}=j/N
\Delta T$ (where $j=1, 2, \ldots, N/2$). 
Thus the highest frequency accessible to the DFT is
the Nyquist frequency, $f_{\rm Nyq}=f_{N/2}=1/2\Delta T$, and the
lowest frequency is $f_{1}=1/N\Delta T$. Note that it  is
customary to subtract the mean flux from the light curve before
calculating the DFT, this eliminates the zero-frequency power.  

The complex valued DFT is then squared
\begin{equation}
\label{eqn:ft2}
|X(f_j)|^2 =
X^{\ast}(f_j) X(f_j) = 
\{ R_X(f_j) \}^2 + \{ I_X(f_j) \}^2,
\end{equation}
where $^{\ast}$ denotes the complex conjugate. 
The periodogram, $P(f_{j})$, is then calculated by choosing an appropriate
normalisation:
\begin{equation}
\label{eqn:pds}
P(f_{j}) = \frac{2}{\bar{x}^{2}} |X(f_j)|^2.
\end{equation}
(Note that the Fourier transform has already been normalised in
equations~\ref{eqn:ft} and \ref{eqn:sine} to account for the data sampling.)

This normalisation (defined in van der Klis 1997, see also Miyamoto
\et 1991) is most commonly used in analysis of AGN and X-ray 
binaries because the integrated periodogram yields the fractional
variance of the data (i.e. $F_{\rm var}^2$). This normalisation will
be used throughout this paper. 
The factor of two is because this is a ``one sided'' normalisation;
the periodogram is defined for both negative and positive frequencies,
with this normalisation the integral over only positive frequencies 
yields the fractional variance.
The integrated periodogram between two
frequencies, say $f_{1}$ and $f_{2}$, yields the contribution to the
fractional variance due to variations between the corresponding
timescales $1/f_{2}$ and $1/f_{1}$ (this follows from Parseval's
theorem, see e.g. van der Klis 1989). The units for the
periodogram ordinate are (rms/mean)$^{2}$ Hz$^{-1}$ (where rms/mean is
the dimensionless quantity $F_{\rm var}$), or simply Hz$^{-1}$.

\subsubsection{Poisson noise}

If the time series is a continuous photon counting signal binned into intervals of
$\Delta T$, as it is here, the effect of Poisson noise is to add an
approximately constant amount of power to the periodogram at all
frequencies. Thus, at high frequencies, where the red noise spectrum
of the AGN is weak, the observed periodogram will be dominated by
the flat (``white'') Poisson noise spectrum.
With the above normalisation this constant Poisson noise
level is $P_N = 2/\bar{x}$. This formally assumes zero background
flux, but for the present observation this is a reasonable
approximation (even in the hard band the background count rate is
$\sim 0.6$ per cent that of the source).

\subsubsection{Scatter in the periodogram}
\label{sect:scatter}

A periodogram measured from a single light curve shows a
great deal of scatter around the underlying PSD.  This is a natural
outcome of measuring the periodogram from a single realisation of a
stochastic process. The result of this is that the periodogram at a
given frequency, $P(f)$, is scattered around the underlying 
PSD, $\mathcal{P}(f)$, following a $\chi^{2}$ distribution with two degrees of
freedom (see Section 6.2 of Priestley 1981):
\begin{equation}
\label{eqn:pds_scatter}
P(f) = \mathcal{P}(f) \chi_{2}^{2}/2,
\end{equation}
where $\chi_{2}^{2}$ is a random variable distributed as $\chi^{2}$
with two degrees of freedom, i.e. an exponential distribution with a
mean and variance of two and four, respectively\footnote{For the case
of even $N$ the Fourier transform at the Nyquist frequency
($DFT(f_{\rm Nyq})$) is always real, and the periodogram at this
frequency  ($P(f_{\rm Nyq})$) is distributed as $\chi_{1}^{2}$, i.e.
with one degree of freedom}. The expectation value of the periodogram
at a frequency $f$ is thus equal to the PSD but so is its standard
deviation, as a result the periodogram shows a great deal of
scatter. This fundamental property of the periodogram is discussed
terms of X-ray time series analysis by e.g. Leahy \et (1983), van der
Klis (1989), Papadakis \& Lawrence (1993), Timmer \& K\"{o}nig (1995)
and  Stella \et (1997).
In addition the periodogram is an {\it inconsistent} estimator of the
PSD because this intrinsic scatter does not decrease as the number of
data points in the light curve is increased. 
The periodogram can, however, be averaged (binned over separate light
curves or adjacent frequencies) to reduce this
scatter and produce a {\it consistent} estimate of the PSD (Jenkins \&
Watts 1968).

In GBHC analysis the light curves are typically broken into many separate
segments and periodograms are calculated for each segment.
At a given Fourier frequency the periodogram estimates
from each light curve are identically and independently
distributed (assuming the variability is stationary) and so their average
will become Gaussian (following the central limit theorem).
Therefore, averaging the periodograms from many independent light
curve segments will produce a consistent estimate of the PSD with
Gaussian errors (e.g. van der
Klis 1997).
However, dividing up the light curve limits the lowest frequency
probed by the periodogram and so is not so suitable for AGN analysis
where it is crucial to examine the widest frequency range possible.

An alternative technique is discussed by Papadakis \& Lawrence (1993)
and is in some sense ``optimal'' for red-noise light curves such as those from
AGN. In this method a single periodogram is computed from the light
curve, reaching down to the lowest accessible frequencies, and
periodogram estimates at consecutive frequencies are averaged.
For evenly sampled
data the periodogram estimates at each Fourier frequency are 
independently distributed (according to
equation~\ref{eqn:pds_scatter}) and so the averaged periodogram estimate in each frequency bin
will tend to a Gaussian distribution.
In the method of Papadakis \& Lawrence (1993) it is 
the logarithm of the periodogram that is binned, such
that each bin contains $N \ge 20$ periodogram estimates, and an error
is assigned based on the scatter within each bin. 
Using the logarithm of the periodogram means the distribution of
the binned periodogram converges on Gaussian with fewer data points
per bin, and therefore allows for finer frequency  binning. This
binning method is used throughout this paper to produce binned
periodogram estimates with errors.  (See also van der Klis
1997 for more on binned periodogram estimates.)

\subsubsection{Bias in the periodogram}
\label{sect:bias}

A further point is that  periodograms measured from finite data tend
to be biased by windowing effects which further complicate their
interpretation (van der Klis 1989; Papadakis \& Lawrence 1993; Uttley
\et 2002). The observed light curve ($f(t)$) is the true, continuous light
curve of the source ($l(t)$) multiplied by the sampling, ``window,''
function ($w(t)$):
\begin{equation}
f(t)=l(t) \cdot w(t),
\end{equation}
where the window function takes the value unity when the light curve is sampled and
zero elsewhere. From the convolution theorem of Fourier transforms,
the Fourier transform of the observed light curve ($F(f)$) is the 
Fourier transform of the true light curve ($L(f)$) convolved with the
Fourier transform of the window function ($W(f)$):
\begin{equation}
F(f)=L(f) \otimes W(f).
\end{equation}
The periodogram (which is the
square of the discrete Fourier transform of the observed light curve)
is therefore the periodogram of the  continuous light curve of the source
multiplied by the periodogram of the window function.
Thus the periodogram distorted away from the true PSD by the sampling, i.e. the convolution
with the spectral window function.

This leads to two main effects. The first is due to the discrete
sampling of the data. If the light curve is not contiguously sampled
then power from frequencies above the Nyquist frequency can be folded
back, or ``aliased,'' into the observed frequency range. For binned,
continuous light curves such as those presented here, the aliasing
will negligible (see van der Klis 1989). In any event, from continuous
data the very high frequency spectrum will dominated by the Poisson
noise level and not aliasing.

The second effect is ``red noise leak'' and arises as a result of the
finite duration of the light curve. Power is transferred from low to
high frequencies by the lobes of the spectral window function (see
e.g. Deeter \& Boynton 1982; van der Klis 1997). If there is
significant power  at frequencies below the lowest frequency probed by
the periodogram (i.e. on timescales longer than the length of the
observation), this can manifest itself as slow rising or falling
trends across the light curve. These trends will contribute to the
total variance of the light curve (and thus the total power) but, as
the periodogram does not extend to such low frequencies, the power
will be transfered into the observed frequency band-pass (i.e. it
will ``leak'' to higher frequencies). Red noise leak tends to add a
constant component to the observed periodogram in the form of a
power-law of slope $\alpha=2$ (i.e. the PSD of a linear
trend; see van der Klis 1997 for more details), 
the amplitude of this depends on the PSD of the source at low
frequencies (see also section 3.3 of Papadakis \& Lawrence 1995). 
The effect of red noise leak on the periodogram can be accounted for using the Monte
Carlo simulation procedure outlined below.

\subsection{PSD fitting procedure}
\label{sect:mc}

The method employed in the present work is similar to the methods
discussed by Done \et (1992), Green, M$^{\rm c}$Hardy \& Done (1999) and Uttley
\et (2002). A model PSD is assumed, multiple realisations  (light
curves) are simulated using this PSD and are resampled to match the original
data (thereby including the distorting effects of the sampling).
Periodograms are then calculated from
the simulated light curves in the same manner as the real data,
averaged over the realisations, and this ``average distorted model'' (ADM), which includes
the effects of red noise leak, is compared to the original
periodogram using a goodness-of-fit measure. 
This method therefore accounts for the biases in the
periodogram introduced by the window function. It is important to bear
in mind that, as with any model-fitting procedure, the results depend
on the assumptions made, i.e. the choice of specific models to test. 

The errors derived from the periodograms (see Section~\ref{sect:scatter})
are approximately Gaussian (Papadakis \& Lawrence 1993) and therefore
the $\chi^{2}$ statistic can be used to estimate the goodness of the
fit. 
Once a reasonable fit is found, uncertainties on the model parameters can be estimated using
the $\Delta \chi^{2}$ criterion (e.g. Lampton, Margon and Bowyer 1976),
as in X-ray energy spectrum fitting. 

The algorithm described by Timmer \& K\"{o}nig (1995) was used to
generate the simulated light curves from the model PSD. This algorithm
generates a random time series from an arbitrary broad-band PSD,  correctly
accounting for the intrinsic scatter in the powers (i.e.
equation~\ref{eqn:pds_scatter}), and is  computationally very
efficient. The effects of red noise leak can be accounted for by
extending the PSD model to much lower frequencies than required, 
generating light curves much longer than the observed light curve and
using only a segment of the required length. Data simulated in
such a fashion will include power on timescales much longer than those
probed by each segment alone.

The fitting procedure is as follows. 

\begin{enumerate}

\item
A model PSD is chosen and 500 random light curves are
generated\footnote{The number of simulated light curves was chosen as
a compromise between the need for an accurate ADM, which requires a
large number of simulations, and processing time. With 500
simulations, the mean $1\sigma$ scatter in each ADM frequency bin is 1
per cent.}  using the algorithm of Timmer \& K\"{o}nig (1995). The
simulated light curves were $\approx 26$ times longer duration than
the orbital light curves in order that red noise leak is accounted for.

\item
From each of the simulated light curves a section from the middle
is resampled to match the data. Specifically, from each (long)
simulated light curve three consecutive light curves were extracted,
to match the three consecutive revolutions, using exactly the same
sampling as the real data.  The logarithmically binned periodogram is
computed in exactly the same way as for the real data
(Section~\ref{sect:pds_firstlook}).

\item
The 500 binned periodograms are then averaged to produce the ADM -
this represents the periodogram after being distorted by the light
curve sampling (folded through the window function).

\item
The constant Poisson noise level is added and the comparison
between the ADM and the data is then made by calculating the
$\chi^{2}$ of the fit. 

\item
The normalisation of the ADM is then adjusted to minimise the
$\chi^{2}$ (keeping the Poisson noise level fixed).

\item
The $\chi^{2}$ of the fit is then evaluated over a grid of PSD models and
the minimum, corresponding to the best-fitting model, is found.

\end{enumerate}

This method differs slightly from the method of Uttley \et (2002).
Firstly, because the \xmm\ light curves are continuous the
effects of aliasing are negligible, thus the simulated light curves
were generated with the same time resolution as the data (100~s). The
sparsely sampled \xte\ data used by Uttley \et (2002) meant that
aliasing was a significant factor and needed to be accounted for in
the simulations. The second major difference is that  Uttley \et
(2002) calculated errors on the ADM (based on the scatter in the
simulated periodograms), and these were used to estimate the goodness
of fit. In the present analysis there are enough data points to apply
the Papadakis \& Lawrence (1993) binning (with $N=20$) and estimate the error on the
observed periodogram  directly from the data, for use with
$\chi^2$-fitting. 

The $\chi^2$ fitting employed in this method relies on the data being
binned sufficiently. It would in principle be possible to make use of
the maximum frequency resolution offered by the unbinned periodogram
by choosing a different goodness-of-fit estimator (e.g. deriving the
rejection probability directly from the Monte Carlo trials, as in
Uttley \et 2002). However, the results (the best-fitting parameters)
are model-dependent in any case and so using a different method is
unlikely to yield any further information.

\subsection{The periodogram of \mcg}
\label{sect:pds_firstlook}

The periodogram of the full-band data was computed by calculating the
periodogram for each revolution of data (following
equation~\ref{eqn:pds}).  The light curves from each revolution, when
taken separately, are perfectly evenly sampled and uninterrupted. This
means that for each light curve the periodogram estimates are
independent at each Fourier frequency, and they are also independent
between the periodograms from the three light curves.  The
periodograms for the three light curves were therefore combined to
produce a single, binned periodogram by sorting the periodogram points
by frequency and then applying the logarithmic binning (Papadakis \&
Lawrence 1993).

Furthermore, no additional windowing was applied to the data (see
Section 13.4 of Press \et 1992 and section 7.5 of Priestley
1981). Windowing is often used to suppress  the leak of power from one
frequency to another, as it suppresses the side-lobes of the spectral
window function $W(f)$ and therefore the amount of power displaced to
distant frequency bins.  However, in exchange for this the central
lobe of the spectral window function is broadened, meaning more power
is lost to neighbouring frequency bins. This transfer of power between
consecutive frequency bins means they are no longer independently
distributed, which is required by the binning method applied above.

\begin{figure}
\rotatebox{270}{
\resizebox{!}{\columnwidth}{\includegraphics{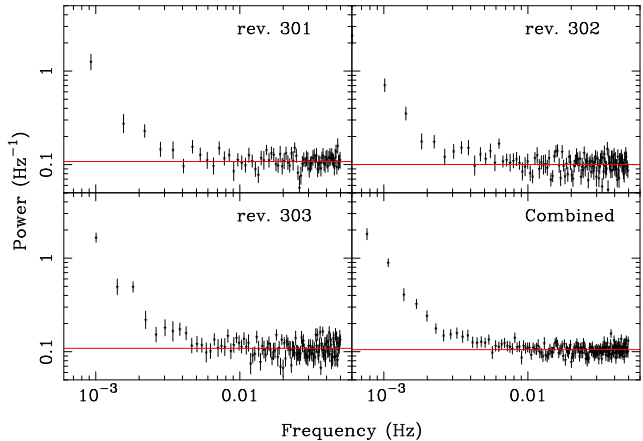}}}
\caption{
High frequency periodograms estimated using 10~s resolution, full-band
light curves.
At high frequencies the periodograms are dominated by the flat (white)
Poisson noise spectrum. The solid lines show the expected Poisson noise levels.
}
\label{fig:highf_pds}
\end{figure}

In order to accurately estimate the Poisson noise level, the
periodograms were first calculated using the full-band light curves with 10~s
resolution. This allows the periodogram to extend up to high
frequencies, at which the variability is dominated by Poisson noise.
Fig.~\ref{fig:highf_pds} shows the periodograms from each of the
three orbits (binned by $N=50$) and the combined periodogram (binned
by $N=100$ to improve the signal-to-noise). The red noise PSD of the
source can be seen below $\sim 3 \times 10^{-3}$~Hz, above which the
(white) Poisson noise spectrum dominates. Also marked are the
predicted noise levels assuming just Poisson noise in the data. These
show that the predicted noise level is clearly an accurate estimate of the
true noise level in the data. In the following analysis only the 100~s
resolution light curves are used, since above the Nyquist frequency
($5\times10^{-3}$~Hz) the periodogram is dominated by the Poisson noise and
not the source variability, and the Poisson noise level is assumed to be 
at the expected value.

\begin{figure}
\rotatebox{270}{
\resizebox{!}{\columnwidth}{\includegraphics{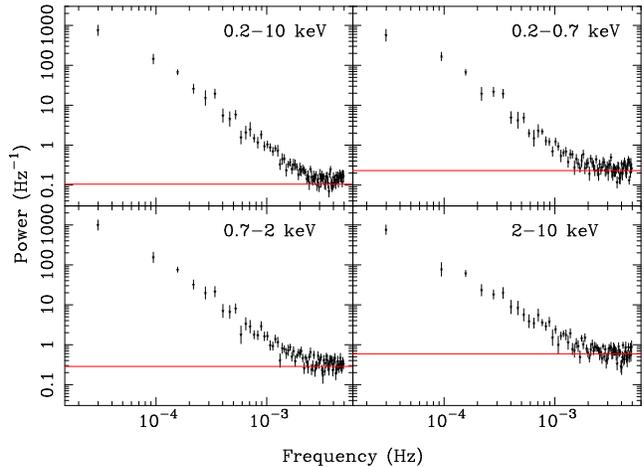}}}
\caption{
Periodograms for each energy band calculated using three
revolutions of data.
The solid lines show the expected Poisson noise levels.
}
\label{fig:dirty_pds}
\end{figure}

Fig.~\ref{fig:dirty_pds} shows the raw periodograms (binned by
$N=20$), calculated by combining the periodograms from each
revolution. The red noise PSD of the source is clearly detected above
the Poisson noise level at frequencies below $\sim$few $10^{-3}$ Hz in
all bands. In the following sections these ``dirty'' periodograms were
fitted using the Monte Carlo procedure in order to derive the properties of
the PSD in a robust manner.

\subsection{Results}
\label{sect:pds_results}

\subsubsection{The full-band PSD of \mcg}

\begin{table}
\centering
\caption{
Results of fitting the full-band periodogram with various trial models.
The models are discussed in the text and shown in Fig.~\ref{fig:models}.
Model 1 is a simple power-law, model 2 is a power-law breaking to a
slope of $\alpha_{\rm low}=1$ at low frequencies, model 3 is a
power-law with a break to $\alpha_{\rm low}=0$, model 4 is a
power-law with a smooth transition to a slope of $\alpha_{\rm low}=1$
at low frequencies and model 5 is an exponentially cut-off power-law.
Errors on the model parameters correspond
to a 90 per cent confidence level for one interesting parameter
(i.e. a $\Delta \chi^{2}=2.7$ criterion).\label{tab:pds_fit1}} 
\begin{center}
\begin{tabular}{lccc}                
\hline
Model      & Slope $\alpha$ & Break freq. $f_{\rm br}$ ($10^{-5}$ Hz) & $\chi^{2}/dof$ \\
\hline
1        & 2.10        &     ---      &  $160.8/79$ \\
2  & $2.50\pm0.15$    & $10_{-4}^{+10}$   &  $87.2/78$  \\
3  & $2.45\pm0.15$ & $5_{-1}^{+3}$  &  $89.0/78$  \\
4  & $2.65\pm0.20$ & $13_{-6}^{+8}$ & $84.1/78$ \\
5  & $1.8\pm0.1$   & $100\pm30$     & $83.4/78$ \\
\hline
\end{tabular}
\end{center}
\end{table}

\begin{figure}
\rotatebox{270}{
\resizebox{!}{\columnwidth}{\includegraphics{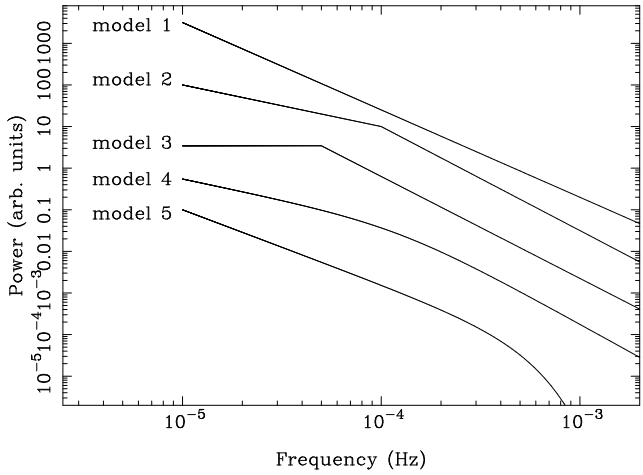}}}
\caption{
Trial PSD models used to fit the full-band data.
}
\label{fig:models}
\end{figure}

\begin{figure}
\rotatebox{270}{
\resizebox{!}{\columnwidth}{\includegraphics{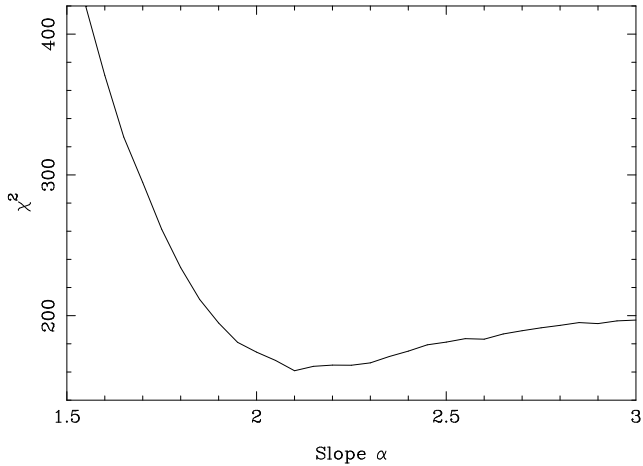}}}
\caption{
Fit statistic against power-law index for a simple power-law PSD model
(model 1, with $79 ~ dof$). 
}
\label{fig:fit_contour1}
\end{figure}

As expected, the periodograms shown in
Fig.~\ref{fig:dirty_pds} show an approximately power-law spectrum
for the intrinsic (source) variability. Various trial
models were fitted to the data (the models are illustrated in
Fig.~\ref{fig:models}), using the procedure outlined above to
constrain the PSD. 
The first trial model was a power-law (model 1):
\begin{equation}
\mathcal{P}(f) = N f^{-\alpha}.
\end{equation}
This model has two free
parameters, namely the slope ($\alpha$) and normalisation ($N$) of the power-law (the
additional constant Poisson noise level, $P_N$, is kept fixed). As summarised in
Table~\ref{tab:pds_fit1}, this model provided an unacceptable fit to the data
($>99.99$ per cent rejection probability). Fig.~\ref{fig:fit_contour1}
shows the variation in $\chi^2$ with $\alpha$ (for 79
degrees of freedom, $dof$). The $\chi^2$
of the fit changes little for PSD slopes steeper than $2$. This is an
effect of the red noise leak. If the PSD continues to very low
frequencies as a steep power-law ($\alpha > 2$), a significant amount
of power will leak into the observed periodogram in the form of an
$\alpha=2$ power-law. This red noise leak contribution will cause the
observed periodogram to have a slope of $\alpha \approx 2$ even when
the true underlying PSD slope is much steeper. Thus, if the
power-law PSD continues unbroken to low frequencies, it is difficult
to distinguish slopes steeper than $2$ in the fitting. 

\begin{figure}
\rotatebox{270}{
\resizebox{!}{\columnwidth}{\includegraphics{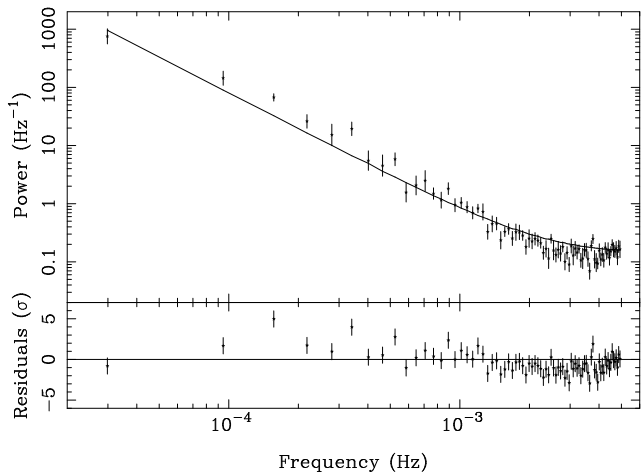}}}
\caption{
The full-band data (points with errors) and best-fitting single
power-law model (model 1; solid line). The residuals from the fit are shown in
the lower panel.
}
\label{fig:model_fit1}
\end{figure}

Fig.~\ref{fig:model_fit1} shows the
best fitting power-law model, with a slope of $\alpha=2.1$.
This steep slope implies that the PSD must break at low frequencies or the
integrated power would diverge. Therefore PSD models including a
flattening at low frequencies were explored in an attempt to constrain
the frequency of the break.
The residuals from the power-law fit show that the model is too flat
above $2 \times 10^{-4}$~Hz and too 
steep at lower frequencies, suggesting the PSD may break in the
observed frequency range.

\begin{figure}
\rotatebox{270}{
\resizebox{!}{\columnwidth}{\includegraphics{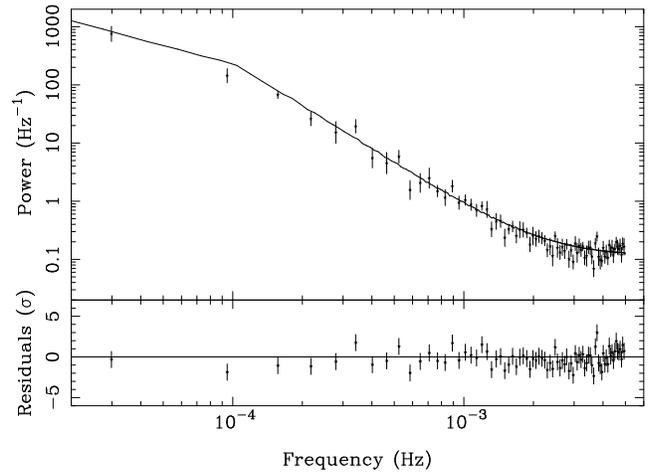}}}
\caption{
The full-band data and (folded) broken power-law model (model 2).
}
\label{fig:model_fit2}
\end{figure}

\begin{figure}
\rotatebox{270}{
\resizebox{!}{\columnwidth}{\includegraphics{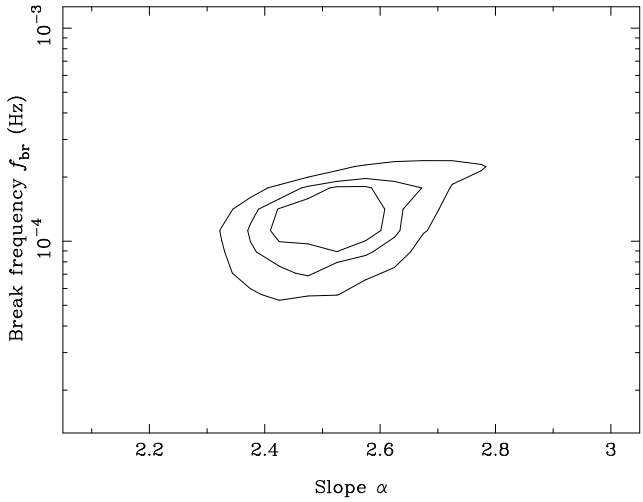}}}
\caption{
Contour plot showing the parameters of model 2. 
The contours represent 68.3, 90 and 99 per cent confidence levels.
}
\label{fig:contour2}
\end{figure}

Indeed, based on long timescale \xte\ monitoring, Uttley \et (2002) found the
best fitting PSD model for \mcg, in the 2--10~keV band, was a broken power-law. When the slope below
the break was fixed at $\alpha_{\rm low}=1$ the slope above the break was
found to be $\alpha=2.0\pm0.3$, with a break frequency in the range
$1.3\times 10^{-5}$ to $1.0\times 10^{-4}$ Hz (90 per cent confidence
limits). Therefore the next trial model fitted was a broken power-law (model 2) 
\begin{equation}
\mathcal{P}(f) = \left\{ \begin{array}  
{l@{\quad:\quad}l}
N \left(\frac{f}{f_{\rm br}}\right)^{-\alpha_{\rm low}} & f \le f_{\rm br} \\
N \left(\frac{f}{f_{\rm br}}\right)^{-\alpha} & f > f_{\rm br},
\end{array} \right.
\end{equation}
with a break from a slope of $\alpha_{\rm low}=1$ to $\alpha$.
The free parameters were the
high frequency slope ($\alpha$), the
normalisation ($N$), and the frequency of the break ($f_{\rm br}$). 
The fit parameters are given in Table~\ref{tab:pds_fit1}. The model
provided an acceptable fit to the data (77 per cent rejection
probability) and the improvement in the fit was $\Delta \chi^{2} =
73.6$ for the addition of one free parameter (the break
frequency), which, according to the $F$-test, is significant at
$>99.99$ per cent confidence. 
Fig.~\ref{fig:contour2} shows the confidence contours for the model parameters and
the fit residuals using this model are shown in Fig.~\ref{fig:model_fit2}.

Although there are only two periodogram points at frequencies below the break,
the detection of the break is nevertheless highly significant. This is
because the high frequency PSD slope is observed to be steeper than $2$, implying
that the effect of red noise leak is not strong. The
steep PSD observed at high frequencies cannot continue below the
lowest observed frequencies;
if the PSD remained steep to low frequencies the contribution from red noise leak
would cause the periodogram to resemble a $\alpha=2$ power-law over the
whole observed frequency range. The steep PSD at high
frequencies must therefore break to a flatter slope close to the lowest
frequencies probed in order that the high frequency periodogram have a steep slope.

\begin{figure}
\rotatebox{270}{
\resizebox{!}{\columnwidth}{\includegraphics{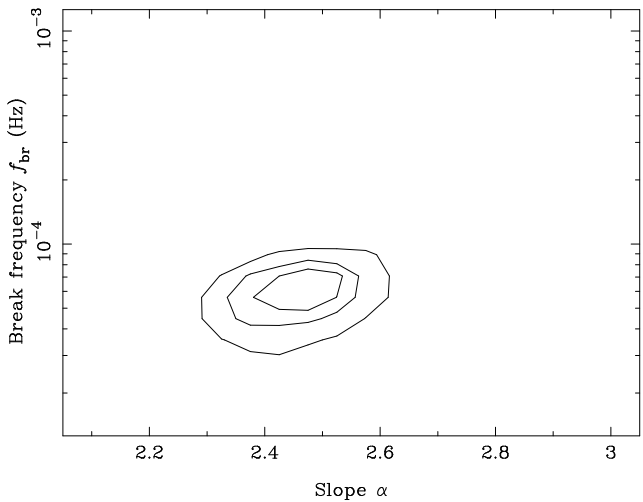}}}
\caption{
As Fig.~\ref{fig:contour2} except for model 3. 
}
\label{fig:contour3}
\end{figure}

The slope of the PSD below the break is not well-constrained from the
\xmm\ data however. Fitting the data with a PSD model that breaks from a
slope of $\alpha_{\rm low} = 0$ (model 3) gives a
similarly acceptable fit but with a slightly lower break frequency
(Table~\ref{tab:pds_fit1}, Fig.~\ref{fig:contour3}).  As these two
models both give good fits it is not possible to distinguish between them
based on these data alone.  
However, it is interesting to note that the break
frequency found from the \xte\ data is consistent with the break
frequency found from the \xmm\ data when the PSD is assumed to
break to $\alpha_{\rm low}=1$. The break frequencies
found from fitting the model assuming a break to $\alpha_{\rm low}=0$
are inconsistent between \xte\ ($f_{\rm br}=5.1_{-2.6}^{+5.1} \times
10^{-6}$~Hz) and \xmm\ ($f_{\rm br}=5_{-1}^{+3} \times 10^{-5}$~Hz). 

\begin{figure}
\rotatebox{270}{
\resizebox{!}{\columnwidth}{\includegraphics{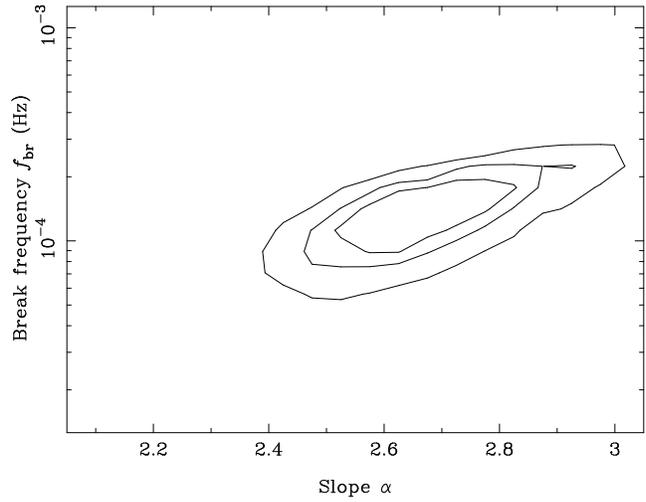}}}
\caption{
As Fig.~\ref{fig:contour2} except for model 4 
}
\label{fig:contour4}
\end{figure}

The exact shape of the break is not well defined from these
data. As an alternative to a sharply broken power-law, a model
representing a power-law with a smooth 
break from a slope of $\alpha_{\rm low}=1$ to $\alpha$ was tested
(model 4):
\begin{equation}
\mathcal{P}(f) = N f^{-1} \left\{ 1 + \left( \frac{f}{f_{\rm br}} \right)^{2} \right\}^{-(\alpha-1)/2}.
\end{equation}
This has three free parameters, $N$, $\alpha$ and $f_{\rm br}$, and
provided an acceptable fit (Table~\ref{tab:pds_fit1},
Fig.~\ref{fig:contour4}), comparable to the models 
including a sharp break, and the break frequency was consistent with
that found for model 2. 

\begin{figure}
\rotatebox{270}{
\resizebox{!}{\columnwidth}{\includegraphics{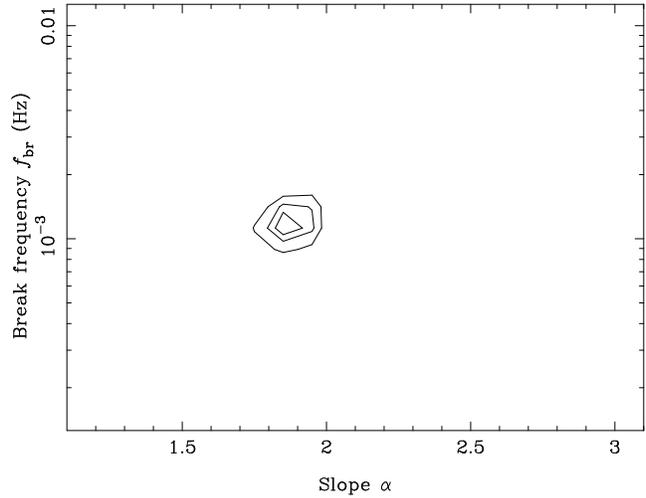}}}
\caption{
As Fig.~\ref{fig:contour2} except for model 5 
}
\label{fig:contour5}
\end{figure}

The final model tested was an exponentially cut-off power-law (model
5): 
\begin{equation}
\mathcal{P}(f) = N f^{-\alpha} \me^{-(f/f_{\rm br})^{2}}
\end{equation}
Again this model has three free parameters, $N$, $\alpha$ and $f_{\rm br}$, and
provided a reasonable fit to the data (Table~\ref{tab:pds_fit1},
Fig.~\ref{fig:contour5}). The best-fitting slope for this model
is much flatter than for the previous models, and the cut-off
frequency is much higher than the break frequencies in the
broken power-law models. That this model provides a good fit
to the data is perhaps suggesting that the PSD of \mcg\ steepens
further at high frequencies.

\subsubsection{Energy dependence of the PSD}
\label{sect:e-psd}

\begin{figure}
\rotatebox{270}{
\resizebox{!}{\columnwidth}{\includegraphics{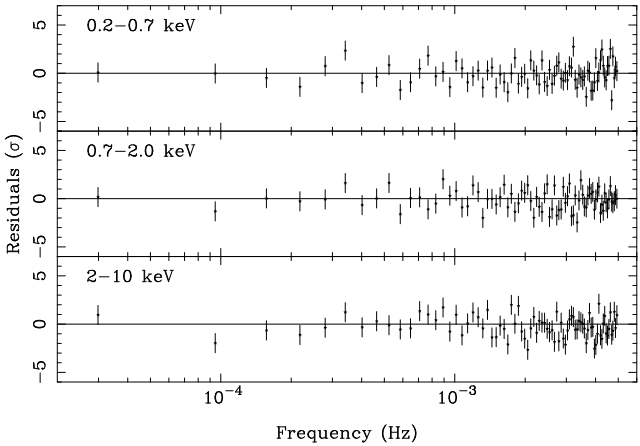}}}
\caption{
Residuals from fits to the soft, medium and hard band data using the
broken power-law model (model 2).
}
\label{fig:pds_fits3}
\end{figure}

\begin{figure}
\rotatebox{270}{
\resizebox{!}{\columnwidth}{\includegraphics{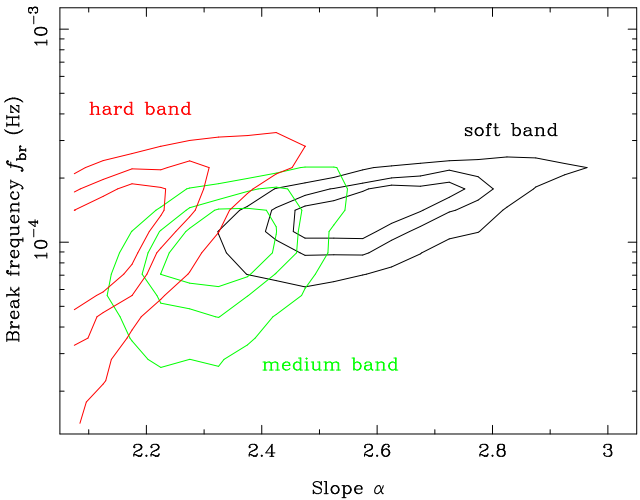}}}
\caption{
Contour plot showing the parameters of the broken power-law model (assuming a
slope of $\alpha_{\rm low}=1$ below the break) for the three energy sub-bands.
The contours represent 68.3, 90 and 99 per cent confidence levels.
}
\label{fig:contour6}
\end{figure}

On the basis of the \xmm\ data alone, PSD models 2 through 5 all
provide reasonable fits to the full-band data.  In the following
analysis, model 2 is fitted  to the data from the soft, medium and
hard energy bands. in order to provide a uniform parameterisation of
the PSD at different energies.  The residuals are shown in
Fig.~\ref{fig:pds_fits3} and the results from these fits are shown in
Fig.~\ref{fig:contour6} and Table~\ref{tab:pds_fit2}.  

The fits to the soft and hard band data using model 2 are somewhat
worse than  the fit to the full band data. However, as can been seen
from Fig.~\ref{fig:pds_fits3}, there are no obvious, strong,
systematic residuals.  Refitting these data with model 4 provided very
similar fits and  refitting using model 5 gave a slightly better fit
to the hard band data but slightly worse fits to the soft and medium
bands. The possibility remains that there is some marginally
significant structure in the PSDs at high frequencies (around $\sim 4
\times 10^{-3}$~Hz in Fig.~\ref{fig:pds_fits3}), not accounted
for in these simple continuum models, as was also
suggested for the high frequency PSD of NGC~7469 by Nandra \&
Papadakis (2001).  These points notwithstanding, model 2 was used to
parameterise, in a simple and uniform fashion, the PSD as a function
of energy.

The break frequencies as determined in each energy band are consistent
with one another. However, the slope of the PSD above the break shows
significant changes between the energy bands, with the hardest band
showing the flattest slope.  Table~\ref{tab:pds_fit2} also gives the
total variability powers, in $F_{\rm var}$ form, derived from each
band (cf. Table~\ref{tab:revs}).  For each band the total power was
estimated by integrating binned periodogram (after subtracting off the
constant Poisson noise level) to give $F_{\rm var, 1}$, and also by
integrating the PSD model (excluding the Poisson noise component) over
the frequency range $10^{-5}$ to $5 \times 10^{-3}$~Hz to give $F_{\rm
var, 2}$. As was noted in Section~\ref{sect:timing}, the medium band
contains the highest total variability power.

\begin{table}
\centering
\caption{
Results of fitting the periodograms derived from the four energy
bands with the broken power-law model (breaking from a slope of
$\alpha$ to slope of $\alpha_{\rm low} = 1$ at low frequencies). 
Errors on the model parameters correspond
to a 90 per cent confidence level for one interesting parameter
(i.e. a $\Delta \chi^{2}=2.7$ criterion).
Also given are the total powers calculated
by integrating the binned periodogram ($F_{\rm var,1}$) and by
integrating the best-fitting PSD model ($F_{\rm var,2}$).
\label{tab:pds_fit2}}
\begin{center}
\begin{tabular}{lccccc}                
\hline
          & Slope          & $f_{\rm br}$ &                & $F_{\rm var,1}$ & $F_{\rm var,2}$ \\
Band      & $\alpha$       & ($10^{-4}$ Hz)           & $\chi^{2}/dof$ & (\%)            & (\%)  \\
\hline
full        & $2.50\pm0.15$ & $10_{-4}^{+10}$  &  $87.2/78$   & 26.0  & 26.3  \\
soft        & $2.55\pm0.20$ & $13_{-5}^{+7}$   &  $108.4/78$  & 23.8  & 23.8  \\
medium      & $2.30\pm0.15$ & $8_{-3}^{+8}$    &  $89.2/78$   & 29.3  & 29.1  \\
hard        & $2.10\pm0.15$ & $10_{-6}^{+10}$  &  $103.5/78$  & 25.4  & 24.0  \\
\hline
\end{tabular}
\end{center}
\end{table}

\subsubsection{Low frequency PSD}

\begin{figure}
\rotatebox{270}{
\resizebox{!}{\columnwidth}{\includegraphics{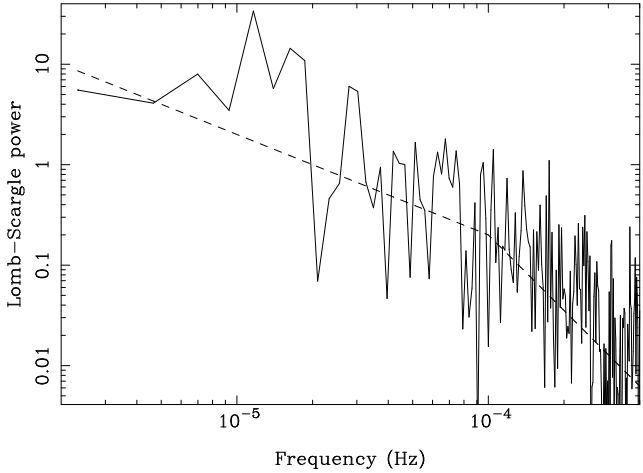}}}
\caption{
Unbinned Lomb-Scargle periodogram of the full-band light curve.
The large scatter in the periodogram is expected 
(see Section~\ref{sect:scatter}).
The dashed line illustrates the broken power-law model used in
Section~\ref{sect:pds_results}. Note that the model has not been
fitted to the data and that the Lomb-Scargle
periodogram is normalised differently from the periodogram discussed
in Section~\ref{sect:periodogram}.
}
\label{fig:lomb}
\end{figure}

The binned periodograms used above reached down to $3
\times 10^{-5}$~Hz in the lowest frequency bin. It would in principle
be possible to treat the three consecutive light curves as a single,
longer light curve (top panel of Fig.~\ref{fig:lightcurves}) and
use this to reach even lower frequencies in the power spectrum. 
However, this three revolution light
curve is no longer uninterrupted, due to the gaps between
revolutions. This means that the standard DFT (as discussed in
Section~\ref{sect:periodogram}) is no longer suitable. There do exist
periodogram estimators designed to deal with data containing gaps,
such as the Lomb-Scargle periodogram (Lomb 1976; Scargle 1982) but
there are drawbacks to using such techniques. Firstly, the gaps in the
sampling mean the periodogram estimates at each frequency are no
longer independent of one another (Scargle 1982), so the binning and error
estimation used above is not strictly valid. Secondly, the more
complicated window function introduces much greater distortion on the
low frequency periodogram, which would severely limit the amount of
information that could be recovered from the lowest frequencies. 
For these reasons the results presented in this paper are based entirely on the
more robust, standard DFT method, but for completeness the
Lomb-Scargle periodogram (computed using the algorithm of Press \&
Rybicki 1989) of the full-band light curve is shown in
Fig.~\ref{fig:lomb}. Down to low frequencies this appears broadly
consistent with the results presented above.  

\subsubsection{Stationarity}

The fundamental assumption underpinning the above analysis is that the
variability process is stationary. A stationary process is one whose
statistical properties do not depend on time (see section 1.3 of
Bendat \& Piersol 1986 for more on stationarity in random
data). Specifically, the assumption made here is that the PSD is
stationary, i.e. that the light curves are all realisations of the
same process. The rms-flux correlation found in Section~\ref{sect:rms-flux}
shows that the variability of \mcg\ is in some sense non-stationary (the variance
does change), but in a ``well-behaved'' fashion. Indeed, the linear
correlation between $\sigma_{\rm rms}$ and $\bar{x}$ means that
the average $F_{\rm var}$ is constant. In other words, the variations, when
normalised by the local flux, appear stationary. 

The method outlined in Appendix A of Papadakis \&
Lawrence (1995) was used  as a simple check of the validity of this
assumption. If the PSD $\mathcal{P}(f)$ is constant then the (unbinned)
periodograms obtained from each revolution should be identical except
for the scatter dictated by equation~\ref{eqn:pds_scatter}. 
Papadakis \& Lawrence (1995) define the statistic $S$ from the
logarithm of the ratio of two
periodograms, summed over a range of frequencies.

\begin{equation}
S = \frac{1}{\sqrt{N}} \sum_{i=1}^{N} 
\frac{\log[P_1(f_i)]-\log[P_2(f_i)]}{\sqrt{{\rm var}\{\log[P_1(f_i)]\}+{\rm var}\{\log[P_2(f_i)]\} }},
\end{equation}
 
where $P_1(f_i)$ and $P_2(f_i)$ are the two periodograms, and ${\rm
  var}\{\log[P_{1,2}(f_i)]\}=0.31$ (Papadakis \& Lawrence 1993).

If the underlying PSD is the same in the two periodograms 
then $S$ should be normally distributed with a mean of zero
and a variance of unity. 

A normalised periodogram was computed for the first 80~ksec of each of
the three
revolutions (the light curves were clipped to equal duration so that
their periodograms had identical Fourier frequencies) and
$S$ was measured for frequencies below $10^{-3}$~Hz, where the source
variability dominates over the Poisson noise. Comparing rev. 301 and
302 gave $S=1.18$, comparing rev. 301 and 303 gave $S=-0.26$
and comparing rev. 302 and 303 gave $S=-1.44$. These three estimates
are all within $2\sigma$ of the expected value assuming stationarity
in the data. Therefore, to first-order at least, the assumption of
stationarity in these data seems a reasonable one.

\section{Cross spectral properties}
\label{sect:cross}

\subsection{The cross spectrum}
\label{sect:cross spectrum}

The PSD is often called the auto spectrum because it
is the Fourier transform of the light curve, multiplied by itself
(equation~\ref{eqn:ft2}). The cross spectrum is a related tool used
for comparing the properties of two simultaneous light curves, such as
from different energy bands, say $x(t)$ and $y(t)$. 
The product of the Fourier transforms of the two light curves gives
the cross spectrum (compare with equation~\ref{eqn:ft2}):
\begin{equation}
\label{eqn:cross_spec}
C(f) = X^{\ast}(f) Y(f).
\end{equation}
In the same way that the PSD (auto spectrum) and
the ACF are Fourier pairs, the cross spectrum and cross correlation
function (CCF) are also Fourier pairs. Thus, mathematically speaking,
the cross spectrum contains the same information as the CCF.

The cross spectrum is complex valued and can be represented
in the complex plane by an amplitude and a phase. The complex Fourier
transforms of the two light curves can be written as $X(f) = |X(f)|
\me^{\mi \phi_x(f)}$ and $Y(f) = |Y(f)| \me^{\mi \phi_y(f)}$ and
the cross spectrum can be written as:
\begin{equation}
\label{eqn:cross_spec2}
C(f) = |X(f)| |Y(f)| \me^{\mi (\phi_y(f) - \phi_x(f))}.
\end{equation}

The phase component of the cross spectrum represents the phase
difference between the two light curves (which corresponds to a time
delay at that frequency) and is discussed more fully in 
Section~\ref{sect:phase}. The squared magnitude of the 
cross spectrum is used to define the ``coherence'' of the two light
curves. 

\subsection{Coherence}
\label{sect:coherence}

\subsubsection{The meaning of coherence}

The coherence, $\gamma^2(f)$,  is a real-valued function and is essentially the
squared cross spectrum normalised by the auto spectra (PSDs) of the
two light curves. 
\begin{equation}
\label{eqn:coherence1}
\gamma^2 (f) = \frac{| \langle C(f) \rangle |^2}{\langle |X(f)|^2 \rangle
  \langle |Y(f)|^2 \rangle}.
\end{equation}
Here the angled brackets represent an averaging over an ensemble of
realisations of the processes.
Coherence differs from the periodogram and time lags in
the sense that it is only meaningful to talk of the coherence from an
ensemble of independent measurements (either from separate light
curve segments or consecutive frequencies). 
Equation~\ref{eqn:coherence1} 
defines the linear correlation coefficient of $X(f)$ and $Y(f)$ (see
Section 9.1 of Priestley 1981)\footnote{The function $\gamma^{2}(f)$
defined by equation~\ref{eqn:coherence1} is often referred to as
the {\it squared
coherence} (Priestley 1981; Bendat \& Piersol 1986) and it gives the square of the linear correlation
coefficient. This function is simply referred to as
the coherence in this paper.}. The 
coherence can thus be thought of as a measure of the degree of linear
correlation between the two light 
curves as a function of Fourier frequency, and is related to the
amplitude of the CCF. 

In order to illustrate coherence, consider two light curves 
that are related by a transfer function $\psi(\tau)$, e.g.
\begin{equation}
\label{eqn:transfer}
y(t)=\int_{-\infty}^{+\infty} \psi(t - \tau) x(\tau) d\tau = \psi(t)
\otimes x(t).
\end{equation}
Then from the convolution theorem 
\begin{equation}
\label{eqn:transfer2}
Y(f) = \Psi(f) X(f),
\end{equation}
where $\Psi(f)$ is the Fourier transform of $\psi(t)$. 
Therefore:
\begin{equation}
|Y(f)|^2 = Y^{\ast}(f)Y(f) = | \Psi(f) |^2 | X(f) |^2,
\end{equation}
and
\begin{equation}
C(f)= X^{\ast}(f) Y(f)  = X^{\ast}(f) \Psi(f) X(f) = \Psi(f) |X(f)|^2.
\end{equation}
The squared magnitude of the cross spectrum then becomes:
\begin{equation}
\label{eqn:transfer4}
|C(f)|^2 = |X(f)|^4 | \Psi(f) |^2.
\end{equation}
And so
\begin{equation}
\frac{ |C(f)|^2 }{ |X(f)|^2 |Y(f)|^2 } = 
\frac{ |X(f)|^4 | \Psi(f) |^2}{ |X(f)|^2 |\Psi(f)|^2 |X(f)|^2 } =
1.
\end{equation}
Therefore the coherence is $\gamma^2(f)=1$. 

This means that if the two light curves are related by a single,
linear transfer function (e.g. a delay or a smoothing) then they will
have unity coherence. In other words,  if knowledge of one light curve
can in principle be used to predict the other, the two are said to be
perfectly coherent.  If, on the other hand, the light curves contain
contributions from completely independent processes then the coherence
will be less than unity.

Another way of thinking about coherence is in terms of
equation~\ref{eqn:cross_spec2}. If the light curves are related by
some simple transform then, at a frequency $f$, there will be a
fixed delay between the light curves, i.e. the lag 
will be the same in each realisation of the process.
If the phase difference is constant then the value of the numerator of
equation~\ref{eqn:coherence1} (the squared mean of
equation~\ref{eqn:cross_spec2}) is equal to the denominator and hence
the the coherence is unity. The interpretation of coherence is
illustrated more fully in  Vaughan \& Nowak (1997) and Nowak \et (1999a).

\subsubsection{Estimating the coherence}
\label{sect:coh_est}

In an analogous manner to the PSD, the cross spectrum is estimated
using the cross periodogram. If the two discretely sampled light
curves, $x(t_i)$ and $y(t_i)$, have DFTs $X(f_j)$ and $Y(f_j)$ then
the cross periodogram is given by the discrete form of
equation~\ref{eqn:cross_spec}, namely:
\begin{equation}
\label{eqn:cross_spec3}
C(f_j) = X^{\ast}(f_j) Y(f_j) = R_C (f_j) + \mi I_C (f_j),
\end{equation}
where $R_C (f_j)$ and $I_C (f_j)$  are the real and imaginary components of
the complex cross spectrum:
\begin{eqnarray}
\label{eqn:cross_comps}
R_C(f_j) = R_X(f_j) R_Y(f_j) + I_X(f_j) I_Y(f_j)
\nonumber \\
I_C(f_j) = R_X(f_j) I_Y(f_j) - I_X(f_j) R_Y(f_j),
\end{eqnarray}
and $R_X(f_j)$ and $I_X(f_j)$ are the real and imaginary components
of the DFT of the light curve $x(t_i)$ as defined by equation~\ref{eqn:sine}.
The coherence is estimated using the
discrete form of equation~\ref{eqn:coherence1}:
\begin{equation}
\label{eqn:coherence2}
\gamma^2 (f_j) = 
\frac{ \langle R_C (f_j) \rangle ^2 + \langle I_C (f_j) \rangle
  ^2}{\langle |X(f_j)|^2 \rangle \langle |Y(f_j)|^2 \rangle}. 
\end{equation}
The angled brackets here represent an averaging over light curve
segments and/or consecutive frequencies.
In practice however, the coherence will be compromised by photon
noise, which is independent between the two light curves and so
will lead to an apparently reduced coherence. As the Poisson noise
power is known, this effect can be removed using the recipe outlined
in Vaughan \& Nowak (1997). In the following analysis the
noise-corrected coherence and its uncertainty were
estimated using equation 8 of Vaughan \& Nowak (1997).
The coherence was calculated only for frequencies below
$10^{-3}$~Hz, above this frequency the Poisson noise power becomes
comparable to the intrinsic source variability and equation~8 of
Vaughan \& Nowak (1997) becomes increasingly inaccurate.

\begin{figure}
\rotatebox{270}{
\resizebox{!}{\columnwidth}{\includegraphics{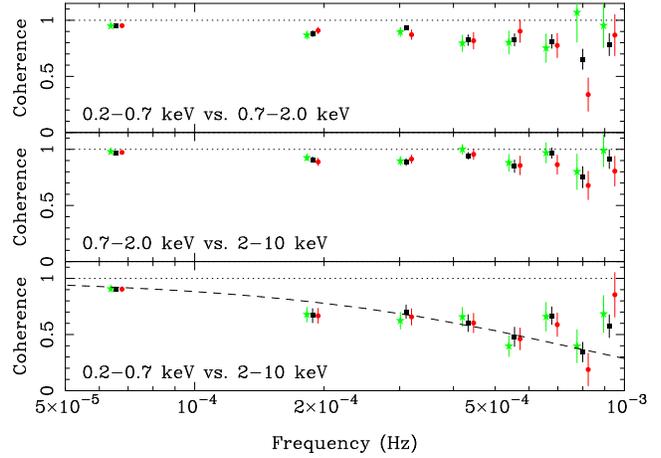}}}
\caption{
Coherence functions between light curves in the three energy bands. The different
symbols mark the results derived using light curves from the three EPIC detectors 
(squares mark pn, circles mark MOS1 and stars mark
MOS2 data). The dotted line shows the function
$\gamma^{2}(f)=\exp(-f/8 \times 10^{-4} ~ \rm{Hz})$ discussed in the text.
}
\label{fig:coherence1}
\end{figure}

\begin{figure}
\rotatebox{270}{
\resizebox{!}{\columnwidth}{\includegraphics{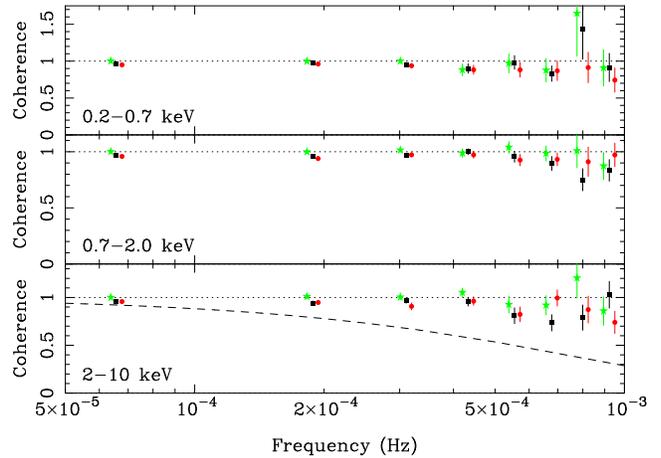}}}
\caption{
Coherence functions between light curves from different detectors in
the same energy bands (squares mark pn vs. MOS1, circles mark pn vs. MOS2 and stars
mark MOS1 vs. MOS2). As these were calculated between light
curves from different detectors but identical energy ranges the
intrinsic coherence is unity and deviations from unity are a result
of Poisson noise. 
The dotted line shows the function plotted in
Fig.~\ref{fig:coherence1} representing the loss of coherence observed
between the energy bands.
Clearly this function shows a much more pronounced loss of coherence than that 
resulting from Poisson noise.
}
\label{fig:coherence2}
\end{figure}

As mentioned
above, the coherence needs to be averaged in some fashion.
For the present analysis the cross  and auto spectra were calculated
separately for each revolution. The data from the
three revolutions were then combined by sorting in order of frequency
and then averaged in bins of $N=40$ frequency points. 

\subsubsection{Coherence properties of \mcg}
\label{sect:coherence_results}

The coherence function was calculated for the three possible combinations of
the three energy bands (soft/medium, soft/hard and medium/hard). This
was repeated using data from MOS1 and MOS2 as well as the pn and the
results were consistent between the detectors. 
(The coherence was also estimated individually for each revolution of
data and the results were consistent between the three revolutions.)
The
resulting coherence functions are shown in
Fig.~\ref{fig:coherence1}. At low frequencies the coherence is close
to unity for all energy bands. At higher frequencies the coherence
becomes much lower, particularly between the soft and hard bands. 
For comparison, Fig.~\ref{fig:coherence1} shows the function
$\gamma^{2} (f) = \exp ( -f/ 8 \times 10^{-4} ~ \rm{Hz})$ compared to the
coherence between soft and hard bands.

This apparent loss of coherence at high frequencies could be an effect
of Poisson noise in the data, since at high frequencies the intrinsic
source power is weaker compared to the noise power. In order to test
this possibility the coherence was evaluated by comparing light curves in
identical energy ranges from the three different detectors. The light curves extracted from
the three detectors should be identical, in a given energy band,
except for Poisson noise. Therefore the coherence should be unity and
any deviation from unity will be a result of Poisson noise. As can
been seen from Fig.~\ref{fig:coherence2}, the coherence is close to unity until the
highest frequencies, and the deviation from unity coherence at high
frequencies is much smaller than the observed loss of coherence
between the different energy bands. 
This, and the fact that the loss of coherence is
repeated in all three detectors suggests that Poisson noise is not seriously
biasing these results, and that the error bars are reasonable
estimates of the true uncertainties. 

\subsubsection{Simulations}
\label{sect:coh_sim}

Simulated data were used as an additional test of the accuracy of the
coherence estimation.  Two simulated light curves were constructed,
representing light curves in two different bands, and the coherence
was calculated for them exactly as for the real data.  These
simulations were used as a further test of the effect of Poisson noise
and to measure the effect of different PSD slopes on the coherence
function.

The effect of Poisson noise was tested by simulating coherent data in
two  bands using the same PSD.  The PSD model used was taken from the
fit to the soft band data using model 2 (Table~\ref{tab:pds_fit2}).  Two
long light curves were generated using this PSD and an identical
random number sequence.   The two light curves, having identical PSDs
as well as Fourier phases and amplitudes,  were thus identical and
therefore have unity coherence at all frequencies.

The simulated light curves for the two bands were resampled to match
the window function of the real data, i.e. three consecutive light
curves were extracted from each to band match the three revolutions of
data.  The light curves for the first band were rescaled to
match the average count rate of the soft band light curve and the
light curves from the second band were rescaled to match the count
rate of the hard band data. 
(The simulated light curves were produced using the absolute PSD
normalisations found from fitting the data. Thus the absolute
variability amplitude of the rescaled simulated data is as expected
based on the real data.)  Poisson noise was then added independently to the
simulated light curves (i.e. the number of counts in each time bin was
randomised according to the Poisson distribution).  These two sets of
simulated data represent soft and hard band light curves in terms of
their Poisson noise, but intrinsically are perfectly coherent and have the
same PSD.

\begin{figure}
\rotatebox{270}{
\resizebox{!}{\columnwidth}{\includegraphics{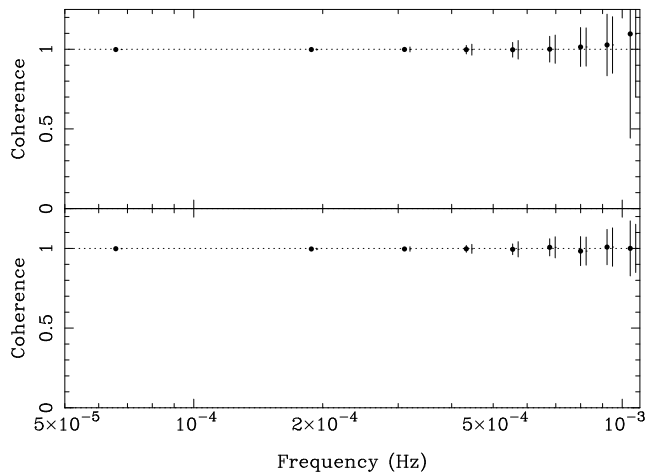}}}
\caption{
Average coherence function measured from (perfectly coherent) simulated data.
The top panel shows results using light curves with
identical PSDs, the bottom panel shows results when different
PSDs are assumed for the two light curves.
The error bars on the data points represent the standard deviation of
the coherence estimates, and the lines without data points represent the average of the
estimated error bars. Clearly the coherence is close to unity and the
error estimates are close to the expected scatter.
}
\label{fig:coherence_sim}
\end{figure}

One hundred sets of simulated data were produced
and for each set the coherence and its error were calculated as
discussed in Section~\ref{sect:coh_est} (i.e. using the recipe of Vaughan
\& Nowak 1997). The average and standard deviation of the coherence
estimates from the 100 sets of simulated data are shown in the top panel of
Fig.~\ref{fig:coherence_sim}. Also shown are the average error bars.
The coherence is clearly very close to
unity over the frequency range examined, and the standard deviation of
the estimates is close to the mean error bar, indicating that the Vaughan
\& Nowak (1997) method did accurately recover the intrinsic coherence
from the effect of the Poisson noise. (Note that these error bars are 
slightly smaller than those derived from the real data because the
uncertainty is itself a function of the intrinsic coherence, which is lower in
the real data.)

The tests above used intrinsically coherent data with identical PSD
shapes. The PSDs fits of Section~\ref{sect:e-psd} show that the PSD
shape is energy dependent. In order to confirm that the coherence is
not biased because of the PSD shape being different in different
energy bands, the
experiment was repeated using PSDs appropriate for the two bands.
Specifically, the best-fitting PSD models for the soft and hard bands
(listed in Table~\ref{tab:pds_fit2}) were used to generate simulated soft
and hard bands light curves, but again the two sets of data used 
an identical
random number sequence. Thus the simulated soft and hard band light
curves again have unity intrinsic coherence but different PSDs.
As before, 100 sets of simulated data were used and the average
coherence and error were calculated. The results are shown in the
bottom panel of Fig.~\ref{fig:coherence_sim}. The average coherence is
again very close to unity, indicating the different PSD slopes in the
different bands did not bias the results.

The loss of coherence observed in \mcg\ is thus not an artifact of
Poisson noise in the data, nor the change in PSD slope with energy, and is repeated
in each of the three EPIC cameras. As such, in the discussion below,
it is assumed that this loss of coherence is intrinsic to the
continuum variations of the source.

\subsection{Time lags}
\label{sect:phase}

\subsubsection{Estimating the time lags}

As mentioned in Section~\ref{sect:cross spectrum}, the cross spectrum
also contains information about the phase differences between the two light
curves (equation~\ref{eqn:cross_spec2}). This information is similar
to the delays measured by the  CCF, except measures the delay as a function of Fourier
frequency. The phase lag is measured from the phase of the averaged
cross periodogram:
\begin{equation}
\label{eqn:phase1}
\phi(f_j) = \arg \left\{ \langle C(f_j) \rangle \right\} = 
\tan^{-1}  \left\{ \frac{\langle I_C (f_j) \rangle}{\langle R_C (f_j) \rangle} \right\},
\end{equation}
where $C(f_j)$ is calculated using equation~\ref{eqn:cross_spec3} and
the angled brackets denote an averaging over consecutive frequency
points and/or light curves.
This phase difference can then be converted into a time delay at
Fourier frequency $f$:
\begin{equation}
\label{eqn:time_lag}
\tau(f_j) = \phi(f_j)/2 \pi f_j .
\end{equation}
Only if the coherence is high at a given frequency is the 
time lag meaningful. If the two light curves are not coherent then
the individual phase estimates from the two light curves are poorly
correlated and so the phase difference of the average cross spectrum has no useful
interpretation. 

For the present analysis, the time lag was calculated using
equations~\ref{eqn:phase1} and \ref{eqn:time_lag}, with the averaging
of the cross spectrum performed in bins of $N=40$ 
frequency points. The error on the time lag estimate was computed
using equation 16 of Nowak \et (1999a). See section 9.1.3 of Bendat \&
Piersol (1986) for a derivation of this equation. 

\subsubsection{Poisson noise effects}
\label{sect:phase_noise}

\begin{figure}
\rotatebox{270}{
\resizebox{!}{\columnwidth}{\includegraphics{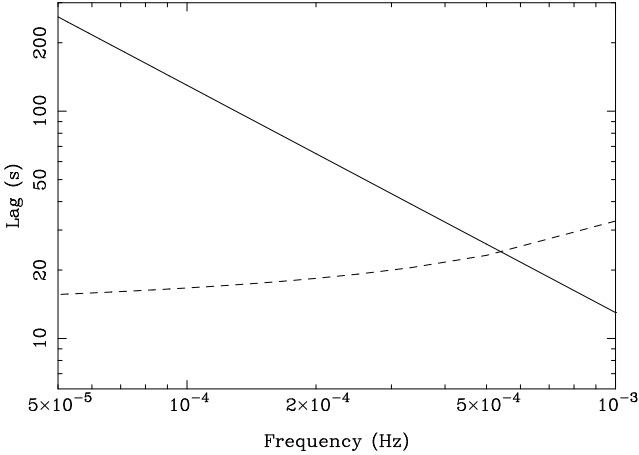}}}
\caption{
Estimate of the sensitivity level of time lag measurements. 
The dotted line shows the uncertainty in the time lag due to Poisson
noise and the solid line represents the approximate time lags
measured, as described by the function given in
Section~\ref{sect:phase_results}. For frequencies above $\approx 6
\times 10^{-4}$~Hz the time lags will not be recoverable from the
Poisson noise.
}
\label{fig:sensitivity}
\end{figure}

The effect of
Poisson noise is to add an independent, random component to the phases of the
two light curves and hence randomise the phase difference. This means
that small time lags are difficult to detect due to photon noise. The
recipe outlined in Nowak \et (1999a) (see their Section 4.2) was used
to estimate the effective noise limit, below which it is difficult to
detect reliable lags. 
The estimated error on the phase lag is computed using their equation
14:
\begin{equation}
\label{eqn:sensitivity}
\Delta \phi (f) \approx \sqrt{ \frac{P_N}{N \gamma^2 (f)}} \frac{1}{\sqrt{\mathcal{P}(f)}}.
\end{equation}
This uncertainty was
calculated using values appropriate for the hard band (PSD
normalisation, slope and noise level $P_N$ from Section~\ref{sect:pds_results}, and
coherence estimated using the function given in
Section~\ref{sect:coherence_results}). This estimated uncertainty,
which should be considered merely a lower limit on uncertainty
introduced by Poisson noise, was compared to the
function used in Section~\ref{sect:phase_results} to illustrate the
observed time lags and the results are shown in
Fig.~\ref{fig:sensitivity}. This indicates that above $\sim 6 \times
10^{-4}$~Hz the lags are too small to be detected, i.e. they fall
below the sensitivity limit defined by the Poisson noise.
A further point is that the coherence is significantly below unity at
higher frequencies ($\gs$few $10^{-4}$~Hz), in which case any measured
time lag is no longer meaningful. Therefore, from these data,
significant time lags will only be detectable below $\sim$few $10^{-4}$~Hz.

\subsubsection{Time lags in \mcg}
\label{sect:phase_results}

\begin{figure}
\rotatebox{270}{
\resizebox{!}{\columnwidth}{\includegraphics{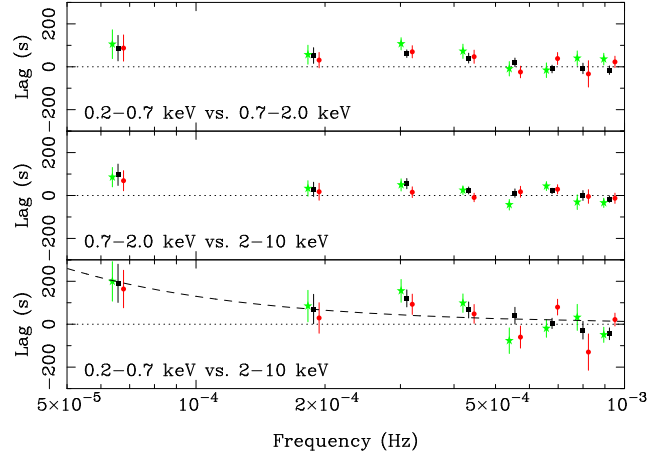}}}
\caption{
Time lags between light curves in the three energy bands. The symbols are as in
Fig.~\ref{fig:coherence1}. A positive lag indicates the first band
leads the second. The dotted line shows the function
$\tau(f)=0.013f^{-1}$ discussed in the text.
}
\label{fig:phase1}
\end{figure}

The time lags were computed (for each EPIC detector) for the three
combinations of energy bands. The results are shown in
Fig.~\ref{fig:phase1}. In each test (three
combinations of energy bands and three different detectors) there was a significant lag
at the lowest frequencies, with the softer band leading the harder band. At
higher frequencies the lag becomes consistent with zero. 
The lag is small, only about $\sim 1$ per cent between soft and hard
bands (e.g. on a timescale of 20,000~s the lag is only $\sim 200$~s).
This is illustrated in Fig.~\ref{fig:phase1}, which shows the
function $\tau(f)=0.013f^{-1}$ compared to the measured time lags
between the soft and hard bands. 

\begin{figure}
\rotatebox{270}{
\resizebox{!}{\columnwidth}{\includegraphics{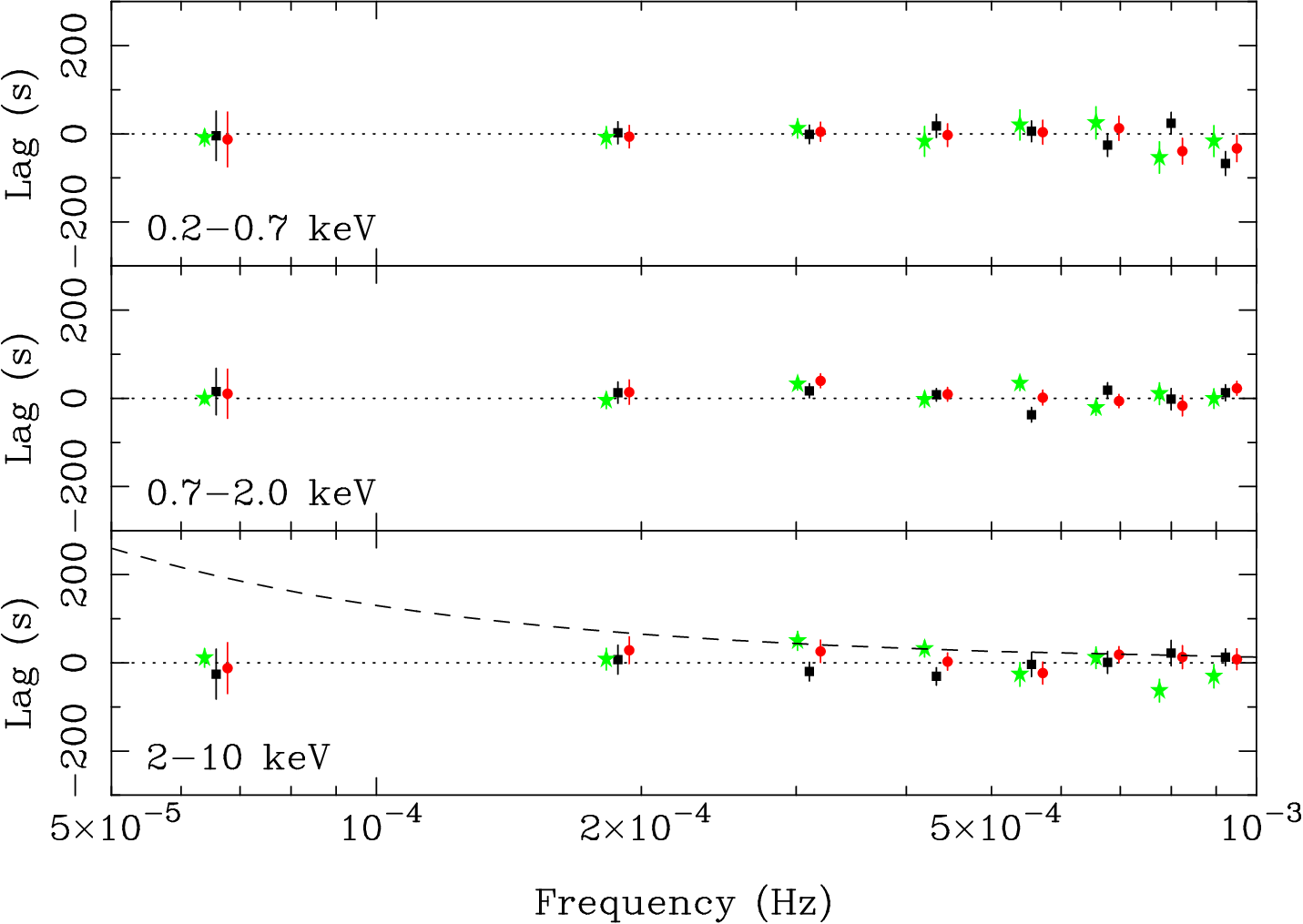}}}
\caption{
Time lags between light curves from different detectors in the same
energy bands. Squares mark pn vs. MOS1, circles mark pn vs. MOS2 and stars
mark MOS1 vs. MOS2.
As the light curves being compared cover identical energy ranges they
should differ only as a result of Poisson noise as so result in zero lag,
any lags observed will be a result of Poisson noise.
The dotted line shows the function plotted in Fig.~\ref{fig:phase1}.
As can be seen, the observed lags between bands
(Fig.~\ref{fig:phase1}) are much larger than those resulting from
Poisson noise.
}
\label{fig:phase2}
\end{figure}

The time lags were calculated using light curves for the three different
detectors in identical energy ranges, to test the possible influence
of Poisson noise (see Section~\ref{sect:coherence_results}). These
are shown in Fig.~\ref{fig:phase2}. In these tests the lag was
consistent with zero over most of the useful frequency range. This,
combined with analysis of Section~\ref{sect:phase_noise} and the fact
that the lags were recovered independently from each EPIC camera, suggests that the lags
observed between different energy bands are not an artifact of Poisson
noise (which in any event would destroy any time lag signal). 

\subsubsection{Simulations}
\label{sect:phase_sim}

The simulated data used in Section~\ref{sect:coh_sim} to assess
the accuracy of the coherence function were also used to assess the
accuracy of the time delay spectrum.

\begin{figure}
\rotatebox{270}{
\resizebox{!}{\columnwidth}{\includegraphics{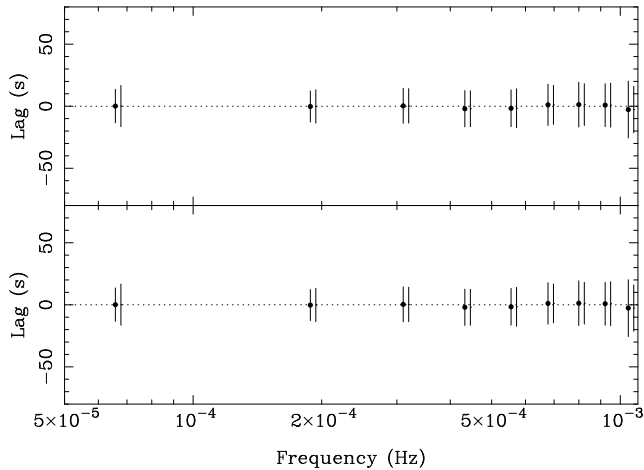}}}
\caption{
Average time delay measured from simulated data (with zero intrinsic delay).
The top panel shows results using light curves with
identical PSDs, the bottom panel shows results when different
PSDs are assumed for the two light curves.
The error bars on the data points represent the standard deviation of
the time delay estimates and the lines without data points represent the average of the
estimated error bars. The phase is close to zero and the
error estimates are close to the expected scatter.
}
\label{fig:phase_sim}
\end{figure}

In the first set of simulations the data in the two bands were
generated using the same PSD and the same random number sequence. In
addition to having unity coherence, the pairs of simulated data also
have identical Fourier phases and therefore zero intrinsic phase lag
between them. The top panel of Fig.~\ref{fig:phase_sim} shows the
average of the time delay estimates, their standard deviation and the
average error estimate from the 100 pairs of simulated data
(including Poisson noise as described in
Section~\ref{sect:coh_sim}). This average spectrum is close to
the zero-delay expectation meaning that the Poisson noise did not
bias the time delay measurements, and the average estimated error is
close to the standard deviation of the estimates.

In the second set of simulations the pairs of simulated data were
generated using different PSDs for the two bands. The bottom panel of
Fig.~\ref{fig:phase_sim} shows the average time delay spectrum,
which again is as expected if the delay estimate is unbiased. It
therefore seems reasonable to assume that the time delays observed in
\mcg\ at low frequencies are intrinsic to the source.

\section{Discussion}
\label{sect:disco}

\subsection{Summary of results}

This paper presents an analysis of the X-ray continuum variability of
\mcg. A strong, linear correlation is found between the rms
variability amplitude and the source flux
(Section~\ref{sect:rms-flux}).  The PSD was examined by fitting model
power spectra to the data after allowing for the distorting effects of
sampling using a Monte Carlo procedure
(Section~\ref{sect:pds}). However, the excellent sampling of the \xmm\
light curves means that these biases are in any case minimised and so
there was a minimum of information lost from the PSD.  The PSD is well
represented by a power-law of slope $\alpha_{\rm low}=1$ at low
frequencies, breaking to a slope of $\alpha \approx 2.5$ at a
frequency $f_{\rm br} \approx 10^{-4}$~Hz. However, it should be
re-iterated that these PSD results are model-dependent, and that
parameterising the PSD using a different model yields slightly
different results.

\begin{figure}
\rotatebox{270}{
\resizebox{!}{\columnwidth}{\includegraphics{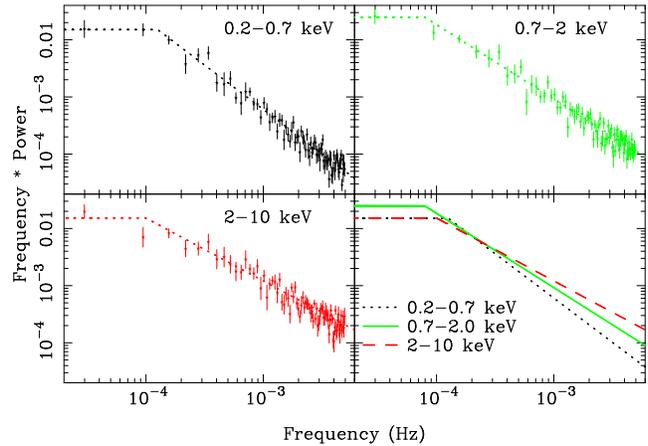}}}
\caption{
Power spectral data from the three energy bands shown in $fP(f)$
units (dimensionless). The data have be ``unfolded'' using model 2 to better
illustrate the shape of the underlying power spectra. The PSD models
are shown with dotted lines, and the error bars show the data/model residuals
multiplied by the unfolded PSD model. The bottom right
panel shows the three model PSDs overlayed.
}
\label{fig:unfolded}
\end{figure}

The PSD is energy dependent, with the variations in the higher energy
band showing a flatter PSD slope (above the break frequency) than
variations in the softer bands. There is no evidence for any energy
dependence of the break frequency. Fig.~\ref{fig:unfolded} illustrates
the PSD shape and energy dependence.  Fitting the PSD as a function of
energy revealed tentative hints of energy-dependent, high-frequency
structure in the PSD (Section~\ref{sect:e-psd}). This could be due to
subtle instrument effects or may represent low-amplitude structure
intrinsic to the source PSD (see also Nandra \& Papadakis 2001).
These timescales ($\gs 10^{-3}$~Hz) are comparable to the dynamical
timescale at innermost stable orbit around the black hole.  However,
further progress on the very high frequency variability of relatively
faint sources such as Seyfert galaxies must await future
high-throughput missions such as \conx\ and \xeus.

The variations observed in different energy sub-bands are highly
coherent at low frequencies (long timescales) but the coherence falls
significantly below unity above $\sim$few $10^{-4}$~Hz, close to the
break in the PSD (Section~\ref{sect:coherence}).  The high coherence
between soft and hard bands at 
low frequencies means that the two bands are linked by a single,
linear transfer function. The direction and magnitude of the time lag
shows that the hard photons lag the soft photons, on average, by $\sim
200$~s at a frequency $f \approx 6 \times 10^{-5}$~Hz
(Section~\ref{sect:phase}). 
It is not
possible to reliably constrain the frequency dependence of the time
lags, but the data seem consistent with an approximately $\tau \propto
f^{-1}$ dependence of the lag (i.e. constant phase difference
$\phi(f)$) similar to that seen in the Seyfert galaxy NGC 7469
(Papadakis \et 2001) and in GBHCs (Nowak \et 1999a; Pottschmidt \et
2000). On shorter
timescales the coherence falls well below unity implying there is no
simple transfer function relating soft and hard bands.

\subsection{Comparison with previous analyses}

Uttley \et (2002) presented an analysis of long timescale monitoring
of \mcg\ using \xte. The PSD was found to be best described by a
broken power-law with slope of $\alpha_{\rm low} = 1$ at low
frequencies, breaking to $\alpha = 2.0\pm 0.3$ at a frequency $f_{\rm
  br} = 0.13 - 1.02 \times 10^{-4}$~Hz.
Using the same PSD model, the break frequency found from the
high-frequency \xmm\ data is $f_{\rm br}=0.6 - 2.0 \times 10^{-4}$~Hz,
consistent with the \xte\ result. The slope of the PSD above the break
is different, but this is most likely an effect of the energy
dependence of the PSD. Indeed, using the 2--10~keV \xmm\ data (i.e. the
same energy range used by Uttley \et 2002) the slope of the high
frequency PSD is $\alpha \approx 2.10\pm0.15$, which is consistent
with the \xte\ result.

The break frequency presented here (and in Uttley \et 2002) is model
dependent. This is because a model PSD had to be assumed and fitted to
the data to derive the appropriate parameters. However, the broken
power-law model (with a break to $\alpha_{\rm low} = 1$ at low
frequencies) provides consistent results when applied to the long
timescale \xte\ monitoring and the short timescale \xmm\ observation. 
This, and the similarity of the observed PSD to those of GBHCs
(see Section~\ref{sect:comparison}) suggests that the chosen model (and
therefore the derived break frequency) is reasonable.

Nowak \& Chiang (2000) used \xte\ and \asca\ data to measure the PSD
of \mcg. They found the PSD to break from $\alpha \approx 0$ to
$\alpha \approx 1$ at a frequency $\sim 10^{-5}$~Hz and then break
again to $\alpha \approx 2$ in the range $10^{-4} - 10^{-3}$~Hz,
The position of the high frequency break, as well as the PSD slopes,
are roughly consistent with those found from the \xmm\ data. However,
as noted by  Uttley \et (2002), the method employed by Nowak \& Chiang
(2000) did not account for distorting
effects of sampling (see Section~\ref{sect:bias}) and underestimated
the errors at low frequencies. Once these effects are included the low-frequency flattening found by
is no longer significant (section~6.2 of Uttley \et 2002).
Nowak \& Chiang (2000) also place a limit on
any time delay between soft (0.5--2.0~keV) and hard (8--15~keV) bands
of $\tau < 2000$~s, consistent with the time lag results presented
here.  

Hayashida \et (1998) also estimated the PSD of \mcg, this time using
\ginga\ data. They found a broken power-law to be a good fit to the
data. However, these authors also did not account for the biases in
the periodogram, which will be
significant due to the patchy sampling of the \ginga\ light
curves. The unusually flat low-frequency slope ($\alpha = -0.5$) found by these authors
is most likely a result of bias in the low-frequency
periodogram, and therefore the position of their claimed break
frequency should be considered with some caution. 

The only other AGN for which detailed cross spectral analysis has been
performed is NGC 7469 (Nandra \& Papadakis 2001; Papadakis \et
2001). In this object the PSD does not show any significant breaks,
but does show the same energy dependence as found in \mcg. The present
analysis shows that the energy dependence of the PSD is due to the slope of
the PSD becoming flatter at higher energies, rather than a change in
the break frequency. In
addition, NGC 7469 shows time lags between energy bands in the same
direction (soft leading hard) and with the similar relative magnitude ($\sim 1$
per cent) as in \mcg. Papadakis \& Lawrence (1995) showed that
in NGC~4051 the harder band showed a flatter PSD than the softer
band. Recently, M$^{\rm c}$Hardy \et (in prep.) used \xmm\ data of NGC
4051 and also found this energy dependence of the PDS.

\subsection{Comparison with Cygnus X-1}
\label{sect:comparison}

Since \exosat\ first showed the medium timescale PSDs of Seyfert
galaxies to have a power-law form, their timing properties have been
compared to those of GBHCs. Now, with long timescale monitoring
available from \xte\ (Edelson \& Nandra 1999; Nandra \& Papadakis
2001; Uttley \et 2002) and short timescale observations possible with
\xmm\ (this paper) it is possible to re-assess this comparison.

The best studied GBHC, Cyg X-1, when in its low/hard state shows a PSD
with an approximately broken power-law form. The PSD breaks from a low
frequency slope of  $\alpha \approx 0$ to $\alpha \approx 1$ and then
to $\alpha \approx 2$ at high frequencies. The position of the high
frequency break is typically $f_{\rm br} \approx 3$~Hz (Belloni \&
Hasinger 1990; Nowak \et 1999a). The comparable break in the PSD of
\mcg\ occurs at $f_{\rm br} \approx 10^{-4}$~Hz. This suggests the
characteristic timescales in these two systems differ by a factor
$\sim 3 \times 10^{4}$. Under the assumption that the timescales scale
linearly with mass of the central black hole, and assuming Cyg X-1 to
contain a $10$~\Msun\ black hole (Herrero \et 1995), then suggests the
mass of the black hole in \mcg\ is $\sim 3 \times 10^{5}$~\Msun\ (in
the range 1.5--5.0 $\times 10^{5}$~\Msun\ using the 90 per cent limits
on the break frequency in \mcg). Assuming a bolometric luminosity of
$4 \times 10^{43}$~erg s$^{-1}$ for \mcg\ (Reynolds \et 1997;
converting to $H_0=75$~km s$^{-1}$ Mpc$^{-1}$) then implies that \mcg\
is radiating at the Eddington luminosity. However, since independent
mass estimates (such as from reverberation mapping) are not available
for \mcg\ it remains possible that this mass estimate is inaccurate if
the relevant timescales do not scale linearly with black hole mass (as
assumed above).

If the frequency of the PSD break is related to the orbital frequency
of matter close to where the peak emissivity of the accretion disc
occurs, $r_{\rm peak}$, then it will depend on $r_{\rm peak}/r_{\rm
in}$, where $r_{\rm in}$ is the inner radius of the disc. For a
standard disc, $r_{\rm peak}/r_{\rm in} \sim 2$. This will also
introduce a dependence on $r_{\rm in}/r_{\rm g}$ (where $r_{\rm
g}=GM/c^2$ is the gravitational radius of the black hole) and thus the
spin parameter of the black hole $a/M$. If the black hole in \mcg\ has
a higher spin parameter than that in Cyg X-1 then the mass deduced
from the above scaling could be increased.

The other temporal characteristics seen in \mcg\ are remarkably
similar to those observed in Cyg X-1, implying there is indeed a
strong connection between these two types of source. The PSD (above
the break frequency) is flatter in harder energy bands in both \mcg\
and Cyg X-1 (Nowak \et 1999a; Lin \et 2000). The strong
rms-flux correlation first seen in X-ray binaries (Uttley \& M$^{\rm
c}$Hardy 2001) is exhibited by \mcg\ (see
Section~\ref{sect:rms-flux}). \mcg\ also displays a loss of coherence
between hard and soft bands at high temporal frequencies (i.e. above
the break in the PSD) very similar to that seen in Cyg X-1 (Nowak \et
1999a) and time lags between these bands that appear similar (in
direction and relative magnitude) to those observed in Cyg X-1.  In
particular, the time lags measured in \mcg\ between soft and hard
bands is $\sim 200$~s in the lowest frequency bin, where the coherence
is near unity. In Cyg X-1 the shortest time lags observed are $\sim 2
\times 10^{-3}$~s (Nowak \et 1999a; Pottschmidt \et 2000) at frequencies just below where
the coherence falls significantly below unity. The ratio of these two
time lags is $\sim 10^5$, similar to the ratio between the PSD break
timescales.

It should also be noted that it is not yet clear whether Seyfert
galaxies more closely resemble the low/hard or high/soft states of
GBHCs (see also the discussion in Uttley \et 2002). In the high/soft
state the PSD of Cyg X-1 is steep at high frequencies ($\alpha \gs 2$)
and flattens to $\alpha \approx 1$ at $f_{\rm br} \approx 10$~Hz, with
no further flattening evident at lower frequencies (Cui \et 1997;
Churazov \& Gilfanov \& Revnivtsev 2001; Reig, Papadakis \& Kylafis
2002). This ratio of characteristic timescales in this case is $\sim
10^5$ and the corresponding mass estimate for \mcg\ is $\sim
10^{6}$~\Msun, implying a luminosity $\sim 30$ per cent of the
Eddington limit.  This high (but nevertheless sub-Eddington) accretion
rate might be expected if \mcg\ is indeed in a state similar to the
high/soft state seen in Cyg X-1. 

It is interesting to note that the best-fitting linear model to the
rms-flux correlation observed in \mcg\ is consistent with passing
through the zero-flux origin (Section~\ref{sect:rms-flux}). The
simplest explanation of this is if there is no significant constant
component to the light curve. A strong constant component ($\sim 25$
per cent of the source flux) is observed from the rms-flux correlation
in Cyg X-1 (Uttley \& M$^{\rm c}$Hardy 2001), but only when the source
is in the low/hard state (Gleissner \et 2002). This is perhaps also
indicating that the variability properties of \mcg\ are more similar
to the high/soft state than the low/hard state of Cyg X-1.
(The time delay properties of Cyg X-1 when in its high/soft state are
rather similar to those seen in its low/hard state, see
e.g. Pottschmidt \et 2000). 

\subsection{Physical implications}
\label{sect:physics}

The X-ray emission mechanism operating in Seyfert galaxies is usually
thought to be inverse-Compton scattering of soft photons in a hot
corona (e.g. Sunyaev \& Titarchuk 1980; Haardt \& Maraschi 1991). In the
simplest models the harder photons are expected to lag behind the
softer photons due to the larger number of scatterings required to
produce harder photons, and the delay should be of order the
light-crossing time of the corona. 
The longest observed time delay in
\mcg\ between soft and hard bands is $\sim 200$~s. Assuming a black
hole mass of $10^6$~\Msun, this lag corresponds to a light-crossing
distance of $\sim 40r_{\rm g}$ ($r_{\rm g}=GM/c^2$). Thus the
direction and magnitude of the observed lag in \mcg\ are consistent
with an origin in a Comptonising corona. However, if the lags are
frequency dependent (as expected by analogy with Cyg X-1 and also seen
in NGC~7469; Papadakis \et 2001) the lags at lower temporal
frequencies would become much longer than expected for a compact
corona (see discussion in Nowak \et 1999b).  
Indeed, the simplest such models predict the
time delay between soft and hard photons to be independent of Fourier
frequency (Miyamoto \et 1988), contrary to what is observed.
In addition, some models
of Compton scattering coronae predict the high frequency PSD should be
steeper for higher energy photons, due to the high-frequency
fluctuations being washed out by multiple scatterings (Hua \&
Titarchuk 1996; Nowak \& Vaughan 1996), again contrary to the observations.
 
Alternatively, the time delay between soft and hard bands could be due
to the spectral evolution of individual X-ray events. If the spectrum
of the event becomes harder as the event unfolds this can lead to an
average lag between soft and hard bands. Such models have been
discussed by B\"{o}ttcher \& Liang (1998) and Poutanen \& Fabian
(1999).

Another alternative is that the delay between soft and hard bands
could be due to a propagation effect, i.e. a ``trigger'' signal
reaching the soft emission region before the hard emission region. In
this scenario the magnitude of the lag is related to the propagation
speed and the distance between soft and hard emission regions. The
following section describes a model based on this kind of signal
propagation.

At higher temporal frequencies the coherence falls well below unity,
meaning there is no longer a simple transfer function between the soft
and hard band emission. As a static corona should maintain a transfer
function (whatever the physical mechanism), on timescales longer than
the break in the PSD ($\gs 10^{4}$~s), where the coherence is high,
the corona may be effectively static (i.e. it maintains its average
properties on these timescales). On shorter timescales the corona may
be dynamic and so the steep PSD above the break and the loss of
coherence may be due to changes in the corona itself.

The timescale of the break in the PSD from $f^{-1}$ to a steeper slope
is $10^{4}$~s, which corresponds to a light-crossing distance of $\sim
2 \times 10^{4} r_{\rm g}$.  The orbital timescale at $40r_{\rm g}$
around a $10^6$~\Msun\ black hole is $t_{\rm orb} \sim 10^3$~s, while
the thermal timescale for a standard, thin accretion disc is $t_{\rm
th} \sim 10^4$~s  (assuming a viscosity parameter $\alpha = 0.1$),
which is of the same order as the timescale of the break in the
PSD. The viscous timescale in such a disc is many orders of magnitude
larger than these other timescales (see the discussion of accretion
disc timescales in section 5.8 of Frank, King \& Raine 1985).
However, if that part of the accretion flow responsible for modulating
the X-ray emission is geometrically thick, the break in the PSD could
correspond to a viscous timescale (which depends on the ratio of disc
height $H$ to radius $R$ as $\sim (H/R)^2$).

\subsection{A phenomenological model}

Various models have been considered to explain the variability of
GBHCs and Seyfert galaxies. These include: various examples of ``shot
noise'' (e.g. Terrell 1972; Lehto 1989; Merloni \& Fabian 2001);
rotating ``hot spots'' on the surface of an accretion disc (Abramowicz
\et 1991; Abramowicz 1992; Bao \& {\O}stgaard 1995); occultation by
moving clouds (Abrassart \& Czerny 2000); self-organised criticality
in an accretion disc (Mineshige, Ouchi \& Nishimori 1994);
fluctuations in the accretion rate propagating through the disc
(Lyubarskii 1997; Churazov \et 2001; Kotov, Churazov \&  Gilfanov
2001) and magnetohydrodynamic instabilities in the inner accretion disc
(Hawley \& Krolik 2001).

Most of these models were developed to explain only one aspect of the
variations (e.g. its red noise nature). Now, however, it is possible
to compare the predictions of these models with the well-determined
timing properties of \mcg. A viable physical model has to be able to
reproduce not only the correct PSD shape and timescales but also the
other temporal characteristics such as the rms-flux correlation, the
energy dependence of the PSD, the time lags and the coherence. It is
not clear whether any of the above mechanism can explain  all the
observations, and indeed some models (e.g. the obscuration model of
Abrassart \& Czerny 2000) seem to offer little hope of explaining
properties such as the coherence and time lags.

A simple phenomenological model has been developed by Lyubarskii
(1997), Churazov \et (2001), and Kotov \et (2001) that seems to
explain (qualitatively, at least) the temporal properties  of Cyg
X-1. As the timing properties of \mcg\ discussed above are very
similar to those of Cyg X-1, it is worth discussing this model as a
possible explanation for the behaviour of Seyfert galaxies.

In its high/soft state, Cyg X-1 shows a $f^{-1}$ PSD over several
decades in frequency (Cui \et 1997; Churazov \et 2001; Reig \et 2002),
down to timescales far longer than would normally be associated with
the X-ray emitting region itself (which is assumed to be compact). The
model put forward by Lyubarskii (1997) to explain this very broad
range in timescale has variations in accretion rate occurring over a
large range of radii from the central black hole and propagating in
towards the X-ray emitting region (assumed to extend over only a few
tens of gravitational radii).

If the viscosity parameter of the accretion flow is varying randomly
and with the same amplitude at all radii, variations in mass accretion
rate at much smaller radii will have an $f^{-1}$ PSD. The accretion rate at
the X-ray emitting region (assumed to be concentrated at small radii)
is thus being modulated by variations produced over a large range in radius.
If the X-ray emission emerging from this 
region is proportional to the local accretion rate then the X-ray
variations will also have an $f^{-1}$ PSD extending down to low frequencies.
Each radius produces variations at a
characteristic timescale comparable to the timescale for inward
propagation, hence long-timescale fluctuations are produced at large
radii and shorter-timescale fluctuations are produced closer in. This
works because any variations on timescales shorter than the
propagation timescale are damped as they move through the disc
(see Lyubarskii 1997, and Churazov \et 2001). This naturally explains both the
$f^{-1}$ part of the  PSD (i.e. below the high frequency break;
Churazov \et 2001) and  the rms-flux correlation (Uttley \& M$^{\rm
c}$Hardy 2001).

Once these variations reach the inner regions, where accretion energy
is released as X-ray emission, variations on shorter timescales are
further suppressed, leading to a steepening of the PSD.  Thus the high
frequency break in the PSD from $f^{-1}$ to a steeper slope (as seen
in \mcg) is due to suppression of fluctuations within the X-ray
emitting region, and the timescale of the break in the PSD corresponds
to the characteristic timescale at the outer radius of the X-ray
emitting region (Churazov \et 2001). The emissivity as a function of
radius defines a Green's function for the incoming $f^{-1}$
fluctuations. In frequency terms, the radial extent of the X-ray
producing region makes it act as a ``low-pass filter'' for the
incident $f^{-1}$ fluctuations, resulting in a break in the PSD, to a
steeper slope, at high frequencies. 

The specific model described by Churazov \et (2001) has a relatively
stable, optically thick, geometrically thin accretion disc sandwiched
by an optically thin, geometrically thick corona. The disc produces
relatively slowly varying thermal emission (which would emerge
presumably in the optical/ultraviolet for a Seyfert galaxy) while it
is the variations within the corona that propagate inwards to produce
the X-ray variability. The timescale of the high frequency break thus
corresponds to the accretion timescale ($\sim$the viscous
timescale) of the geometrically thick corona at the outer edge of the
X-ray emitting region.

Kotov \et (2001) extended this model by adding the simple condition
that the spectrum of the X-ray emitting region is a function of
radius, with the outer regions producing a softer spectrum than the
inner regions. This model can then reproduce the observed time lags
between soft and hard emission. Accretion rate variations propagating
inwards reach the softer emitting region (larger radii) before the
harder emitting region (smaller radii), explaining why the soft
emission leads the hard. This can explain the energy dependence of the
time lags (as delay increases with the separation between energy
bands) as well as the frequency dependence (since different radii
correspond to different characteristic frequencies as well as
different energy spectra). In addition, this model explains the energy
dependence of the PSD above the high frequency break. The harder
emission is produced from smaller radii, which can be modulated by
high frequency fluctuations, hence the harder emission will have more
high frequency power than the softer emission (i.e. the hard band PSD
will be flatter), as observed (see Fig.~\ref{fig:unfolded}). Also, as
the variations on short timescales are occurring within the X-ray
emitting region, i.e. are a result of the dynamic corona, the
coherence between bands will fall off at high frequencies, as mentioned in
Section~\ref{sect:physics} (see also the discussion in Nandra \&
Papadakis 2001). 

Such phenomenological models require that variations occurring at large radii can
be fed into much smaller radii and so influence the PSD. The observed
iron line profile of \mcg\ from this observation (Fabian \et 2002)
indicates that the disc extends down to less than $2r_{\rm g}$, and, from
the steep inferred emissivity profile, that much of the X-ray power is
generated very close to the black hole. This presents a challenge for
all such models, given that the dynamical timescale of this innermost
region is $10$--$100$~s, whereas the timescales measured (e.g. the PSD
break timescale) are orders of magnitude longer. If the
$f^{-1}$ PSD of \mcg\ is similar to that in the high/soft state of Cyg
X-1 (Reig \et 2002) then the variability timescales will extend to
many orders of magnitude longer still. Indeed, the total power in
$f^{-1}$ spectra diverges so slowly that the PSD could in principle
continue as $f^{-1}$ until very low frequencies without the total
power becoming large. (For example, model 2 from
Section~\ref{sect:pds_results} could continue down to the unphysically
low frequency $\sim 10^{-28}$~Hz without the integrated power exceeding $F_{\rm var}=1$!)
 
\section{Conclusions}
\label{sect:conc}

This paper presents a detailed time series analysis of the continuum
variability of \mcg. The PSD shows a break from $f^{-1}$ to $f^{-2.5}$
at a frequency $f_{\rm br} \sim 10^{-4}$~Hz. Comparing the break
timescale with the analogous timescale seen in Cyg X-1, and assuming
that these scale linearly with black hole mass, gives a black hole
mass of $\sim 10^6$~\Msun\ for \mcg\ and suggests it is accreting at a
significant fraction of the Eddington limit. The slope of the PSD
above the break is energy dependent, with the harder band showing more
high frequency power than the softer band. The variations in different
energy bands also become incoherent above the break in the PSD. This
leads us to speculate that on timescales shorter than the break
timescale, the X-ray emitting corona is dynamic, and that harder
photons are produced  at smaller radii  than softer photons.  At the
lowest Fourier frequencies the soft and hard bands are highly coherent
and the soft band is found to lead the hard band by $\sim
200$~s. The direction and magnitude of this lag is consistent with
simple Comptonisation models. A phenomenological model, originally
developed to explain the $f^{-1}$ PSD of GBHCs such as Cyg X-1, is
discussed and found to be able to reproduce (qualitatively) the
observed temporal characteristics of \mcg.  Although this model is
rather simple and phenomenological it can explain the observed
temporal characteristics of the X-ray continuum in \mcg.

\section*{Acknowledgements}

Based on observations obtains with \xmm, an ESA science mission with
instruments and contributions directly funded by ESA Member States and
the USA (NASA). We are grateful to Phil Uttley for many valuable discussions on 
power spectral issues and to Mike Nowak for discussions on coherence estimation.
SV acknowledges many valuable discussions with P. Thomas and financial
support from PPARC. We thank an anonymous referee for useful comments.

\bsp
\label{lastpage}

\begin{thebibliography}{}
\bibitem{1} Abramowicz M. A., Bao G., Lanza A., Zhang X.-H., 1991, A\&A, 245, 454
\bibitem{1a} Abramowicz M. A., 1992, in S. Holt, S. G. Neff, C. M. Urry eds. {\it  Testing the AGN paradigm}, p69 
\bibitem{1b} Abrassart A., Czerny B., 2000, A\&A, 356, 475
\bibitem{2} Belloni T., Hasinger G., 1990, A\&A, 227, L33
\bibitem{3} Belloni T., Psaltis D., van der Klis M., 2002, ApJ, 572, 392
\bibitem{4} Bendat J. S., Piersol A. G., 1986, {\it Random Data: Analysis and Measurement Procedures}, Wiley (New York)
\bibitem{5} Bevington P. R., Robinson D. K., 1992, {\it Data Reduction and Error Analysis for the Physical Sciences}, McGraw-Hill (New York)
\bibitem{6} Bloomfield P., 2000, {\it Fourier Analysis of Time Series}, Wiley (New York)
\bibitem{6f} Bao G., {\O}stgaard E., 1995, ApJ, 443, 54
\bibitem{7} B\"{o}ttcher M., Liang E. P., 1998, ApJ, 506, 281
\bibitem{7f} Brandt W. N., Boller Th., Fabian A. C., Ruszkowski M., 1999, MNRAS, 303, L58
\bibitem{8} Churazov E., Gilfanov M., Revnivtsev M., 2001, A\&A,  321, 759 
\bibitem{9} Cui W., Zhang S. N., Focke W., Swank J. H., ApJ, 484, 383
\bibitem{10} Deeter J. E., Boynton P. E., 1982, ApJ, 261, 337
\bibitem{14} Done C., Madejski G. M., Mushotzky R. F., Turner T. J., Koyama K., Kunieda H., 1992, ApJ, 400, 138
\bibitem{16} Edelson R., Nandra K., 1999, ApJ, 514, 682
\bibitem{18} Edelson R., Turner T. J., Pounds K. A., Vaughan S., Markowitz A., Marshall H., Dobbie P., Warwick R. S., 2002, ApJ, 568, 610
\bibitem{20} Fabian A. C., 1979, Proc. R. Soc. London, Ser. A, 366, 449
\bibitem{21} Fabian A. C. \et 2002, MNRAS, 335, L1
\bibitem{21e} Frank J., King A., Raine D., 1985, {\it Accretion Power in Astrophysics}, Cambridge Univ. Press (Cambridge)
\bibitem{21f} Gleissner T., Wilms J., Pottschmidt K., Uttley P., Nowak M. A., Staubert R., 2002, in Ph. Durouchoux, Y. Fuchs and J. Rodriguez eds. {\it Proceedings of the 4th Microquasar Workshop}, in press (astro-ph/0207610)
\bibitem{22} Green A. R., M$^{\rm c}$Hardy I. M., Lehto H. J., 1993, MNRAS, 265, 664
\bibitem{23} Green A. R., M$^{\rm c}$Hardy I. M., Done C., 1999, MNRAS, 305, 309
\bibitem{23f} Haardt F., Maraschi L., 1991, ApJ, 380, L51
\bibitem{23h} Hawley J. F., Krolik J. H., 2001, ApJ, 548, 348
\bibitem{23j} Herrero A., Kudritzki R. P., Gabler R., Vilchez J. M., Gabler A., 1995, A\&A, 297, 556
\bibitem{23k} Hayashida K., Miyamoto S., Kitamoto S., Negoro H., Inoue H., 1998, ApJ, 500, 642
\bibitem{24} Jansen F. \et 2001, A\&A, 365, L1 
\bibitem{24d} Jenkins G. M., Watts, D. G., 1968, {\it Spectral Analysis and its Applications}, Holden-Day (San Fancisco)
\bibitem{24f} Kotov O., Churazov E., Gilfanov M., 2001, MNRAS, 327, 799
\bibitem{25} Lampton M., Margon B., Bowyer S., 1976, ApJ, 208, 177
\bibitem{26} Lawrence A., Papadakis I., 1993, ApJ, 414, L85
\bibitem{27} Leahy D. A., Darbro W., Elsner R. F., Weisskopf M. C., Kahn S., Sutherland P. G., Grindlay J. E., 1983, ApJ, 266, 160
\bibitem{28} Lehto H. J., 1989, in J. Hunt, B. Battrick, eds, {\it Two Topics in X Ray Astronomy}, (ESA SP-296; Noordwijk: ESA), p499
\bibitem{28c} Lin D., Smith I. A., B\"{o}ttcher M., Liang E. P., 2000, ApJ, 531, 963
\bibitem{28d} Lomb N. R., 1976, Ap\&SS, 39, 447
\bibitem{28f} Lyubarskii Y. E.\ 1997, MNRAS, 292, 679
\bibitem{28h} Markowitz A. \et 2002, ApJ, submitted
\bibitem{28i} M$^{\rm c}$Hardy I. M., 1989, in J. Hunt, B. Battrick, eds, {\it Two Topics in X Ray Astronomy}, (ESA SP-296; Noordwijk: ESA), p1111
\bibitem{28k} Merloni A., Fabian A. C., 2001, MNRAS, 328, 958
\bibitem{29} Mineshige S., Ouchi N. B., Nishimori H., 1994, PASJ, 46, 97
\bibitem{29d} Miyamoto S., Kitamoto S., Mitsua K., Dotani T., 1998, Nature, 336, 450
\bibitem{29f} Miyamoto S., Kitamoto S., 1989, Nature, 342, 773
\bibitem{30} Miyamoto S., Kimura K., Kitamoto S., Dotani T.,  Ebisawa K., 1991, ApJ, 383, 784
\bibitem{33} Mushotzky R. F., Done C., Pounds K. A., 1993, ARA\&A,  31, 717
\bibitem{34} Nandra K., Papadakis I. E., 2001, ApJ, 554, 710
\bibitem{35} Nowak M. A., Vaughan B. A., 1996, MNRAS, 280, 227
\bibitem{36} Nowak M. A., Vaughan B. A., Wilms J., Dove J. B., Begelman M. C., 1999a, ApJ, 510, 874
\bibitem{36b} Nowak M. A., Wilms J., Vaughan B. A., Dove J. B., Begelman M. C., 1999b, ApJ, 515, 726
\bibitem{37} Nowak M. A., Chiang J., 2000, ApJ, 531, L13
\bibitem{37f} Nowak M. A., Wilms J., Dove J. B., 2002, MNRAS, 332, 856 
\bibitem{38} Papadakis I. E., Lawrence A., 1993, MNRAS, 261, 612
\bibitem{39} Papadakis I. E., Lawrence A., 1995, MNRAS, 272, 161
\bibitem{40} Papadakis I. E., Nandra K., Kazanas D., 2001, ApJ, 554, L133
\bibitem{40a} Poutanen J., Fabian A. C., 1999, MNRAS, 306, L31
\bibitem{40f} Pottschmidt K., Wilms J., Nowak M. A., Heindl W. A., Smith D. M., Staubert R., 2000, A\&A, 357, L17
\bibitem{41} Press W. H.\ 1978, Comments on Astrophysics, 7, 103
\bibitem{41a} Press W. H., Rybicki G. B., 1989, ApJ, 338, 277
\bibitem{42} Press W. H., Teukolsky S. A., Vetterling W. T., Flannery B. P., 1992, {\it Numerical Recipes}, Cambridge Univ. Press. (Cambridge)
\bibitem{46} Priestley M. B., 1981, {\it Spectral Analysis and Time Series}, Academic Press (London)
\bibitem{46a} Rees M. J., 1984, ARA\&A, 22, 471
\bibitem{46b} Reig P., Papadakis I., Kylafis N. D., 2002, A\&A, 383, 292
\bibitem{47} Reynolds C. S., Ward M. J., Fabian A. C., Celotti A., 1997, MNRAS, 291, 403
\bibitem{27f} Scargle J. D., 1982, ApJ, 263, 835
\bibitem{48} Stella L., Arlandi E., Tagliaferri G., Israel G. L., 1997, in T. Subba Rao, M. B. Priestley, O. Lessi eds., {\it Applications of Time Series Analysis in Astronomy and Meteorology}, Chapman \& Hall (London) (astro-ph/9411050)
\bibitem{49} Str\"{u}der L. \et 2001, A\&A, 365, L18 
\bibitem{49d} Sunyaev R. A., Titarchuk L. G., 1980, A\&A, 86, 121
\bibitem{49f} Terrell N. J., 1972, ApJ, 174, L35
\bibitem{49h} Thorne K. S., 1974, ApJ, 191, 507
\bibitem{50} Timmer J., K\"{o}nig M., 1995, A\&A, 300, 707
\bibitem{56} Uttley P., M$^{\rm c}$Hardy I. M., 2001, MNRAS, 323, 26
\bibitem{58} Uttley P., M$^{\rm c}$Hardy I. M., Papadakis I., 2002, MNRAS, 332, 231
\bibitem{60} van der Klis M., 1989, in  H. Ogelman, E. P. J. van den Heuvel eds., {\it Timing Neutron Stars}, Kluwer (Dordrecht), NATO ASI Series C 262, p27
\bibitem{62} van der Klis M., 1995, in  W. H. G. Lewin, J. van Paradijs, E. P. J. van den Heuvel eds., {\it X-ray Binaries}, Cambridge Univ. Press (Cambridge), p252
\bibitem{64} van der Klis M., 1997, in  G.J. Babu, E. D. Feigelson eds., {\it Statistical Challenges in Modern Astronomy II}, Springer-Verlag (New York), p321
\bibitem{66} Vaughan B. A., Nowak M. A., 1997, ApJ, 474, L43
\end{thebibliography}
\end{document}